\documentclass[a4paper,11pt]{article}
\pdfoutput=1 % if your are submitting a pdflatex (i.e. if you have
\usepackage{jheppub} % for details on the use of the package, please
\usepackage[T1]{fontenc} % if needed
\usepackage[utf8]{inputenc}

\usepackage[english]{babel}    
\usepackage{xspace}
\usepackage{microtype}    
\usepackage[utf8]{inputenc}    % Allows for writing special charachters in the tex-file 

\usepackage[T1]{fontenc}
\usepackage[extramarks]{titleps}
\usepackage{lipsum}

\usepackage{amsmath,amssymb,amsfonts,amscd,dsfont,bm}     % Standard mathematics
\usepackage{amsthm}
\allowdisplaybreaks     %Allows pagebreak during multiline math enviroments
\usepackage{mathrsfs}
\usepackage[dvipsnames]{xcolor}
\usepackage{slashed}
\usepackage{youngtab}
\usepackage{physics}
\usepackage{orcidlink}
\usepackage[capitalise]{cleveref}
\allowdisplaybreaks

\usepackage{natbib}
\usepackage{graphicx}
\usepackage{hyperref}
\usepackage{comment}

\crefformat{equation}{Eq.~(#2#1#3)}
\crefformat{table}{Tab.~(#2#1#3)}
\crefformat{figure}{Fig.~(#2#1#3)}
\crefformat{appendix}{App.~(#2#1#3)}
\crefformat{section}{Sec.~(#2#1#3)}
\crefformat{subsection}{Subsec.~(#2#1#3)}

\usepackage[normalem]{ulem}

\title{
\huge
Non-uniqueness of boundary-value problems in Renormalization Group flows
}

\author[a]{Astrid Eichhorn\,\orcidlink{0000-0003-4458-1495}} 
\author[a]{Zois Gyftopoulos\,\orcidlink{0009-0007-9439-9284}}
\author[b]{Aaron Held\,\orcidlink{0000-0003-2701-9361}}
\affiliation[a]{
Institut de Physique Théorique Philippe Meyer, Laboratoire de Physique de l’\'Ecole normale sup\'erieure (ENS), Universit\'e PSL, CNRS, Sorbonne Universit\'e, Universit\'e Paris Cité, F-75005 Paris, France
}
\emailAdd{aaron.held@phys.ens.fr}

\affiliation[a]{Institut f\"ur Theoretische Physik, Universit\"at Heidelberg, Philosophenweg 12 \& 16, 69120 Heidelberg, Germany}

\emailAdd{eichhorn@thphys.uni-heidelberg.de}

\abstract{The Renormalization Group flow connects microscopic to macroscopic descriptions of a system and is therefore typically considered as an initial-value problem.
Motivated by situations in which different couplings within a system of Renormalization Group equations are constrained at different scales, we instead consider boundary-value
problems in Renormalization Group flows. We find that, unlike initial-value problems which provide $n$ conditions for $n$ couplings, boundary-value problems which provide $n$ conditions for $n$ couplings do not always have a unique solution. When the Jacobian matrix, i.e., the matrix of first derivatives of beta functions, has complex eigenvalues, boundary-value problems may be non-unique.
We provide a diagnostic tool for non-uniqueness in systems with many couplings. We also provide two examples with potential relevance for physics, namely within the Standard Model as well as within the Einstein-Hilbert truncation of asymptotically safe quantum gravity.}

\begin{document}
\setcounter{tocdepth}{2}
%\notoc{\notocfalse}
\maketitle

%\newpage
%=======================================================================================================

%%%%%%%%%%%%%%%%%%%%%%%%%%%%%%%%%%%%%%%%%%%%%%%%%%%%%%%%%
\section{Introduction and motivation}
\label{sec:intro}
%%%%%%%%%%%%%%%%%%%%%%%%%%%%%%%%%%%%%%%%%%%%%%%%%%%%%%%%%
Renormalization Group (RG) flows relate physics at different scales by evolving couplings in an (effective) Lagrangian. The scale dependence is encoded in a set of beta functions for the scale-dependent couplings $\vec{g}(t)$. Here, $\vec{g}$ is a vector that summarizes all couplings in the system. By $t$ we denote the RG ``time'', which is a dimensionless measure of scale, constructed from the logarithm of the ratio of the RG scale to a reference scale $t=\ln \left(\mu/\mu_0\right)$. There are numerous ways in which beta functions can be calculated, and subtle differences in the meaning of different forms of RG running, see \cite{Buccio:2024hys} for an interesting example. These differences are not important for us; we instead focus on an overarching aspect of RG flows that applies generally. 

Traditionally, RG flows are initialized in the ultraviolet (UV) and evolved towards the infrared (IR), realizing the idea that macroscopic, effective descriptions of a physical system emerge from microscopic ones. In other settings, e.g., within the Standard Model of particle physics, one often inverts the problem; providing initial conditions at IR scales and evolving towards the UV, e.g., \cite{Buttazzo:2013uya}, to discover whether there is a scale of new physics that is indicated by a breakdown of this evolution, e.g., \cite{Hambye:1996wb,Bezrukov:2012sa}.
In terms of the system of coupled ordinary differential equations, first order in RG ``time'', the mathematical problem that is solved in these cases is an \emph{initial-value problem} (IVP).\footnote{The Renormalization Group forms a semigroup, where, upon integrating from the UV to the IR, microscopic information is encoded in IR-irrelevant operators, so that the UV could only be recovered upon knowing infinitely many higher-order couplings with arbitrarily high precision. Nevertheless, a finite set of beta functions simply constitutes a set of coupled ordinary differential equations, for which initial conditions can be supplied in the UV or the IR, i.e., formally, the flow from the IR to the UV can also be considered in such a system.} 
This IVP is guaranteed to have a unique solution, provided the beta functions are sufficiently smooth. Put differently, RG trajectories cannot cross and therefore, specifying $n$ initial conditions for $n$ couplings uniquely specifies a trajectory.

There are, however, settings, in which the values of distinct couplings are fixed at distinct scales. A first example of this is in experimental determinations of couplings, where distinct experiments determine distinct couplings and operate at different energies. Examples can be found, e.g., in \cite{ParticleDataGroup:2024cfk} for the Standard Model of particle physics where the masses of different particles are determined through experiments of different characteristic energy scales and translated into values of couplings at a common reference scale.
We assume here that the dependence on physical energy scales is tracked by the dependence of the coupling on the RG scale (see, e.g., \cite{Buccio:2023lzo,Buccio:2024hys} for discussions as well as cautionary counterexamples). A second example is a setting in which a theoretical determination of the value of a particular coupling comes out of a UV completion or UV extension of a given system, while the other couplings in the system are determined by experiments at lower scales. One specific case for this second class of examples is the asymptotically safe Standard Model, reviewed in \cite{Eichhorn:2022jqj,Eichhorn:2022gku}, in which a UV completion of the Standard Model with gravity determines the values of some, but not all couplings at the Planck scale \cite{Shaposhnikov:2009pv, Eichhorn:2017ylw,Eichhorn:2017lry,Eichhorn:2018whv, Eichhorn:2025sux}.

In both examples, the RG trajectories are determined by a \emph{boundary-value problem} (BVP). At a first glance, one may be tempted to conclude that the uniqueness of the IVP guarantees uniqueness of the BVP. However, this is not in general the case in the theory of coupled ordinary differential equations \cite{keller2018numerical}. In fact, existence and uniqueness of local solutions have only been established in certain BVPs \cite{Peterson1977,Agarwal1986_TwoPointProblems}, and to the best of our knowledge, such a theory remains incomplete. Thus, we are led to wonder: Under which circumstances does specifying $N$ conditions for $N$ couplings, with $k$ conditions at some UV scale $k_{\rm UV}$ and $N-k$ conditions at some IR scale $k_{\rm IR}$, uniquely select a single RG trajectory?\\

This paper is structured as follows: In Sec.~\ref{sec:construction} we explain that BVPs for the RG flow can be non-unique and explain the origin of such non-uniqueness. In Sec.~\ref{sec:fp-examples} we construct explicit two-dimensional examples, in which non-uniqueness hinges on complex eigenvalues of the Jacobian matrix, i.e., the matrix of first derivatives of the beta functions. As a particular example, we consider the Einstein-Hilbert sector of asymptotically safe quantum gravity and show that the BVP can be non-unique in this system, unless the cosmological constant is small enough, as it is in our universe. In Sec.~\ref{sec:SM} we provide a diagnostic tool to detect non-uniqueness in a statistical test and show that there is non-uniqueness in a BVP for the RG flow between the Planck and the electroweak scale in a two-generation subsector of the Standard Model. We also consider Renormalization Group improved settings in a gravitational theory.
We conclude and provide an outlook on future directions in Sec.~\ref{sec:conclusions}.

%%%%%%%%%%%%%%%%%%%%%%%%%%%%%%%%%%%%%%%%%%%%%%%%%%%%%%%%%
\section{Construction of non-unique boundary value problems}
\label{sec:construction}
%%%%%%%%%%%%%%%%%%%%%%%%%%%%%%%%%%%%%%%%%%%%%%%%%%%%%%%%%
%
\begin{figure*}[!t]
    \begin{centering}
        \hfill
        \includegraphics[width=0.8\linewidth]{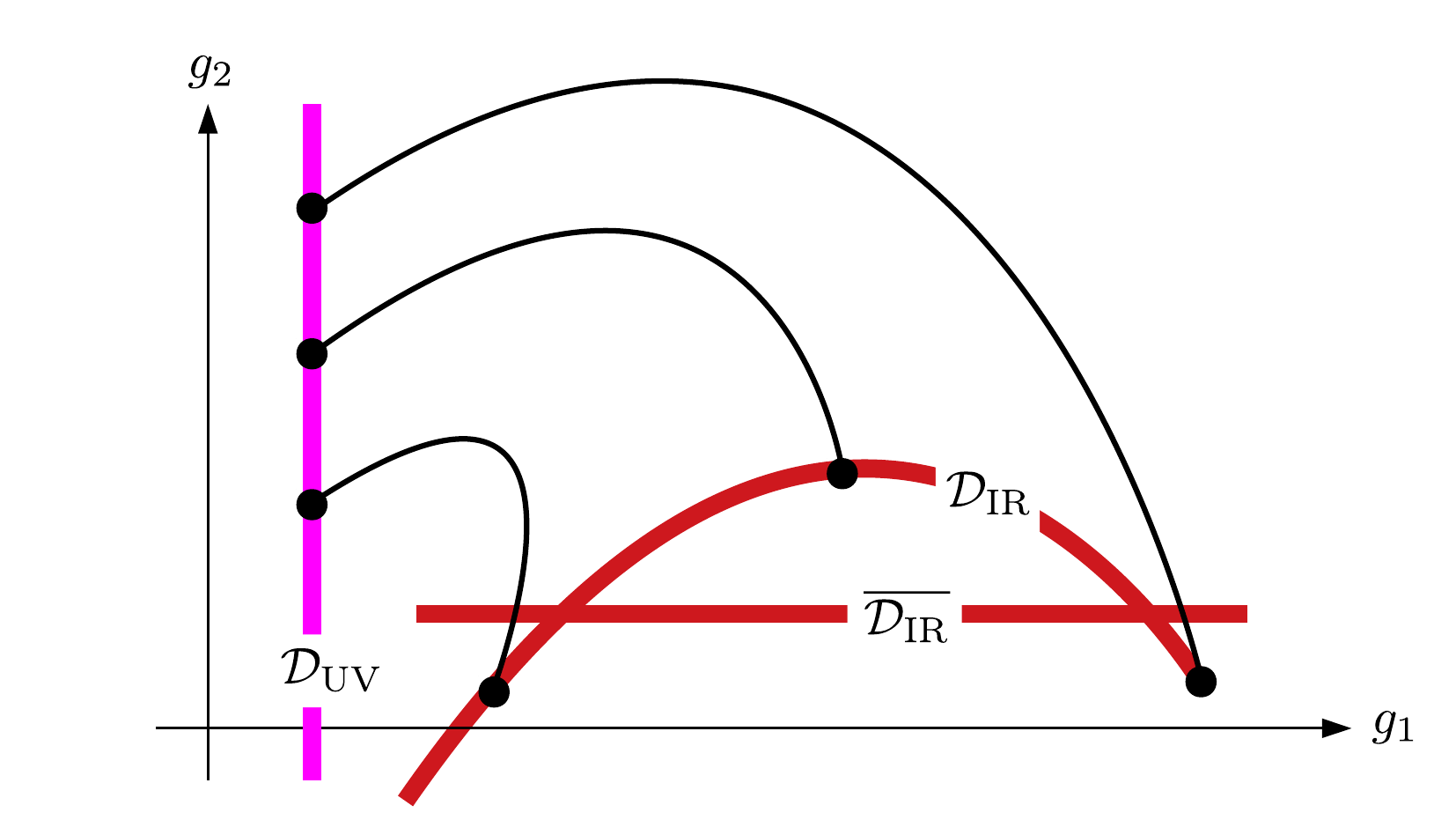} 
        \hfill        
    \end{centering}
    \caption{\label{fig:BVP}
    Sketch of a situation in which the BVP on $\mathcal{D}_\text{UV}=(g_2(t_\text{UV})\equiv a)$ and $\mathcal{D}_\text{IR}$ (infinitely many solutions) also implies non-uniqueness of the BVP on $\mathcal{D}_\text{UV}=(g_2(t_\text{UV})\equiv a)$ and $\overline{\mathcal{D}_\text{IR}}=(g_1(t_\text{IR})\equiv b)$ (finitely many solutions).
    }
\end{figure*}

In general, BVPs are non-unique, if the surface $\mathcal{D}_{\rm IR}$ that results from evolving the UV boundary surface $\mathcal{D}_{\rm UV}$ intersects the IR boundary surface $\overline{\mathcal{D}_{\rm IR}}$ more than once, see Fig.~\ref{fig:BVP}.
To formalize this diagnostic, we consider a set of $N$ beta functions
\begin{align}
    \partial_t\mathbf{g}=\beta_g\;,
    \quad
    \mathbf{g}\in\mathbb{R}^N,
\end{align}
describing the evolution of couplings in RG ``time'' $t$.
The following construction determines whether or not there is a non-unique BVP. (The construction may seem almost trivial, but will still be helpful once we apply it to special cases of interest.)
\begin{itemize}
    \item Start with an $N-k$ dimensional boundary $\mathcal{D}_\text{UV}$ at the UV scale $t=t_\text{UV}$.
    \item Assuming that the IVP is globally well posed (no divergences or other obstructions), the RG evolution of initial data on $\mathcal{D}_\text{UV}$ from $t=t_\text{UV}$ to $t=t_\text{IR}$ produces an IR hypersurface $\mathcal{D}_\text{IR}$. 
    Even if $\mathcal{D}_\text{UV}$ is not curved, $\mathcal{D}_\text{IR}$ is generally curved, i.e., not parallel to any of the axes in the space of couplings. This surface depends on $t_{\rm UV}- t_{\rm IR}$.
    \item By construction, the BVP that is determined by demanding $\mathcal{D}_\text{UV}$ at $t_\text{UV}$ and $\mathcal{D}_\text{IR}$ at $t_\text{IR}$ thus has infinitely many solutions. This is a trivial consequence of considering $N$ couplings, but only specifying $k$ initial conditions, resulting in a $N-k$ dimensional boundary surface, instead of a single initial point.
    This, however, is not in general the BVP that one is interested in, because $\mathcal{D}_\text{IR}$ is not in general a hypersurface that one has specified a priori.
    \item The IR boundary of interest is determined by another surface $\overline{\mathcal{D}_\text{IR}}$, which we consider to generically be flat, i.e., parallel to some of the axes in our space of couplings, and therefore specified by the condition $g_i(t_{\rm IR}) = c_i$, for some $i$ and some constants $c_i$. Altogether, the $N-k$ dimensional UV boundary surface and $k$ dimensional IR boundary surface add up to $N$ conditions on $N$ couplings.
    \item If $\overline{\mathcal{D}_\text{IR}}$ and $\mathcal{D}_\text{IR}$ intersect more than once, then the BVP specified by the UV surface $\mathcal{D}_{\rm UV}$ at $t=t_{\rm UV}$ and the IR surface
$\overline{\mathcal{D}_\text{IR}}$ at $t=t_{\rm IR}$ has more than one solution.
\end{itemize}

To consider an explicit example (cf.~\cref{fig:BVP}), consider $N=2$ and $k=N-1=1$ and start with a straight line, e.g., $\mathcal{D}_\text{UV}=(g_2(t_\text{UV})\equiv a)$. The boundary $\mathcal{D}_\text{IR}$ is also a line ($k=1$), but generally not a straight one. However, we can intersect $\mathcal{D}_\text{IR}$ with a straight line, e.g., $\overline{\mathcal{D}_\text{IR}}=(g_1(t_\text{IR})\equiv b)$. If there are two (or more) intersection points $\overline{\mathcal{D}_\text{IR}}\cap\mathcal{D}_\text{IR}$, these are the values at which the BVP defined by $(g_2(t_\text{UV})\equiv a,\,g_1(t_\text{IR})\equiv b)$ is non-unique.

Continuity implies that the above BVP is non-unique whenever we find three points $A$, $B$, $C$ on $\mathcal{D}_\text{IR}$ such that (i) $g_2(A)<g_2(B)<g_2(C)$ and further (ii) $g_1(A),g_1(C)< b$ while $g_1(B)> b$ or vice versa. Once such points are found, it should be possible to use gradient descent methods to iteratively approximate the respective solutions.

%%%%%%%%%%%%%%%%%%%%%%%%%%%%%%%%%%%%%%%%%%%%%%%%%%%%%%%%%
\section{Two-dimensional examples: complex eigenvalues of the Jacobian matrix}
\label{sec:fp-examples}
%%%%%%%%%%%%%%%%%%%%%%%%%%%%%%%%%%%%%%%%%%%%%%%%%%%%%%%%
%
In what follows, we discuss the situation of a two-dimensional space of couplings. Our results should have a straightforward generalization to higher dimensions (see~\cref{sec:gauge-Yukawa-example} for an example), but we focus on two dimensions because this setting can be visualized clearly and is the lowest dimension in which non-uniqueness is possible.

We discuss toy examples that do not -- to the best of our knowledge -- correspond to the beta functions of any concrete system, but are rather just polynomial expressions which we build to develop an understanding of the possible origin of non-uniqueness of the BVP. We start from a toy example in which non-uniqueness is built in through an ``exotic'' scenario, namely an RG limit cycle. We then adjust the RG flows step by step towards more generic settings, while keeping non-uniqueness. This allows us to build intuition about the prerequisites for non-uniqueness.

\begin{figure}[!t]
\begin{center}
\includegraphics[width=0.49\linewidth]{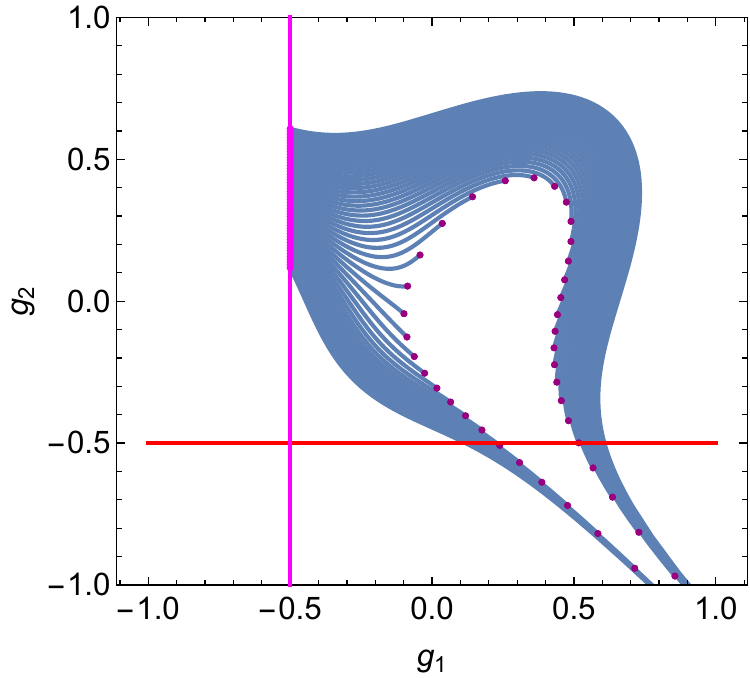}\quad \includegraphics[width=0.45\linewidth]{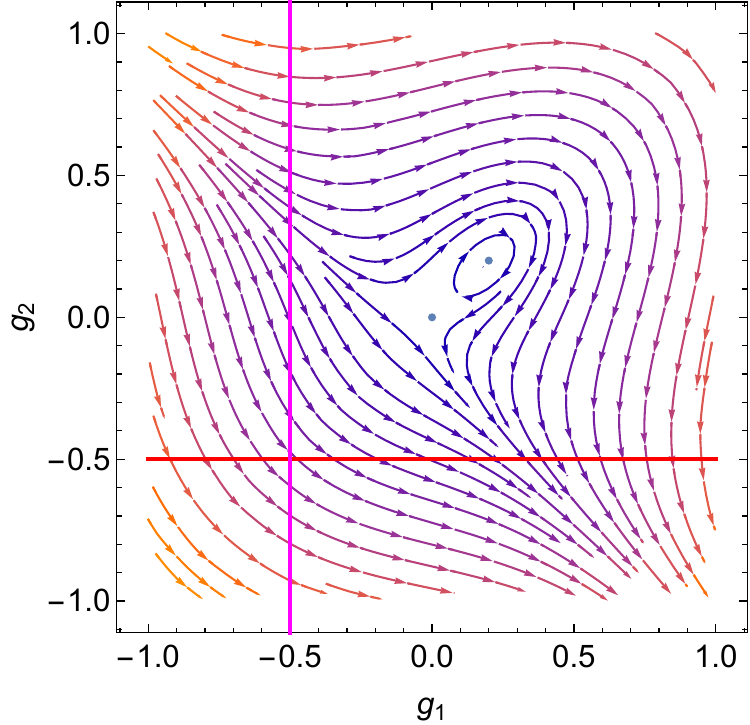}
\end{center}
\caption{\label{fig:onionplot} Left panel: We show a set of RG trajectories, all starting with the initial condition $g_1(k_{\rm UV})=-1/2$ (magenta line). The endpoints of the trajectories (purple points) mark the value when the same amount of RG ``time'' has elapsed on each of them. One can see that there are two trajectories that end exactly on the red line, corresponding to our final condition $g_2(k_{\rm IR}) = -1/2$.
Right panel: We show the corresponding RG flow where colors from blue to yellow indicate growing velocity of the flow defined by Eq.~\eqref{eq:beta1} and \eqref{eq:beta2}. There is a Gaussian fixed point, in addition to an interacting fixed point. The latter is surrounded by a limit cycle.}
\end{figure}

The BVP can have non-unique solutions in the following setting, discussed in general dimensions above and shown in Fig.~\ref{fig:BVP}. In the UV, we start RG trajectories that satisfy a boundary condition that fixes the value of one of the couplings. We evolve the RG trajectories starting from the points along this initial-value line for a fixed amount of RG time, which creates a curve of endpoints of trajectories.
If this curve intersects the line that defines the second IR boundary condition more than once, the BVP has multiple solutions. Clearly, the number of intersection points gives us the exact number of solutions. An example is shown in Fig.~\ref{fig:onionplot}. The example is based on the idea that to have multiple trajectories connecting two boundaries within the same amount of RG time, the flow should ``curve'' and trajectories that curve more should be slower. This suggests a setup with a limit cycle, e.g., the beta functions
\begin{eqnarray}
\beta_{g_1} &=& 2 g_1 - 10\, g_2^2,\label{eq:beta1}\\
\beta_{g_2}&=& -2 g_2 + 10\, g_1^2.\label{eq:beta2}
\end{eqnarray}
The respective flow is plotted in Fig.~\ref{fig:onionplot} and admits two fixed points, $(g_{1,\,\ast},g_{2,\,\ast})=(0,0)$ and $(g_{1,\,\ast},g_{2,\,\ast})=(0.2,0.2)$. The second of these fixed points is actually contained within a limit cycle, see Fig.~\ref{fig:onionplot}. The RG flow between the UV boundary and the IR boundary can go around the region containing the two fixed points in two different ways. To be specific, we call them the ``left-turning'' and the ``right-turning'' trajectories, because they have to ``take a left/right turn'' around the fixed-point region. Because there is a fixed point in the center that slows the RG flow, there are pairs of equal-duration trajectories which contain one left-turning and one right-turning trajectory each. 

The origin of non-uniqueness in this system is clearly correlated with the limit cycle, which enables two sets of trajectories reach the IR boundary condition within the same amount of RG time. We note that the limit cycle gives rise to imaginary critical exponents that characterize the fixed point. This constitutes our first hint that complex eigenvalues of the Jacobian matrix,
\begin{equation}
\mathcal{J} = \frac{\partial \beta_{g_i}}{\partial g_j},
\end{equation}
may lie at the heart of non-uniqueness of the BVP.

However, the limit cycle occurring for the system of beta functions in Eq.~\eqref{eq:beta1} and \eqref{eq:beta2} is not a generic situation for RG flows. In fact, limit cycles are under debate as to whether they should be accepted as possible features of physically viable RG flows \cite{Bulycheva:2014twa,Jepsen:2020czw}, even though they are realized, e.g., in systems exhibiting bound-state formation through the Efimov effect \cite{LeClair:2002ux,Braaten:2004rn}. The reason for this debate is that they constitute apparent violations of the $a$- and $c$-theorems \cite{Zamolodchikov:1986gt,Cardy:1988cwa,Anselmi:1997am,Komargodski:2011vj}. It has, however, been pointed out that these theorems are not violated by RG flows with limit cycles, as long as the quantity that should decrease under the RG flow is multi-valued and can transition from one Riemann sheet to the next under the RG flow around the limit cycle \cite{Curtright:2011qg}. 

To work our way towards a general understanding of the mechanisms underlying non-uniqueness of the BVP, we present a number of paradigmatic examples, first replacing the limit cycle by fixed points, and eventually considering examples without fixed points.

All of the investigated examples are consistent with the hypothesis that the non-uniqueness of a BVP is related to an intermediate flow with rotational behavior, i.e., a Jacobian matrix with complex eigenvalues
\begin{equation}
{\rm Im}\left({\rm eig}(\mathcal{J})\right) \neq 0\;,
\end{equation}
somewhere along the flow that connects the respective boundaries.

%%%%%%
\subsection{Case I: Degenerate fixed point and its deformations}
%%%%%%

\begin{figure}[!t]
\includegraphics[width=0.45\linewidth]{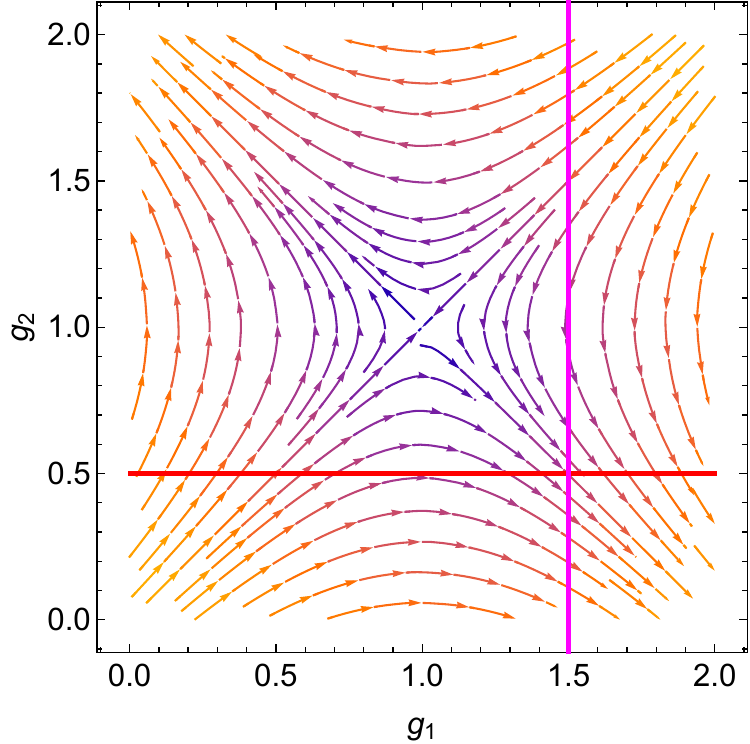}\quad \includegraphics[width=0.45\linewidth]{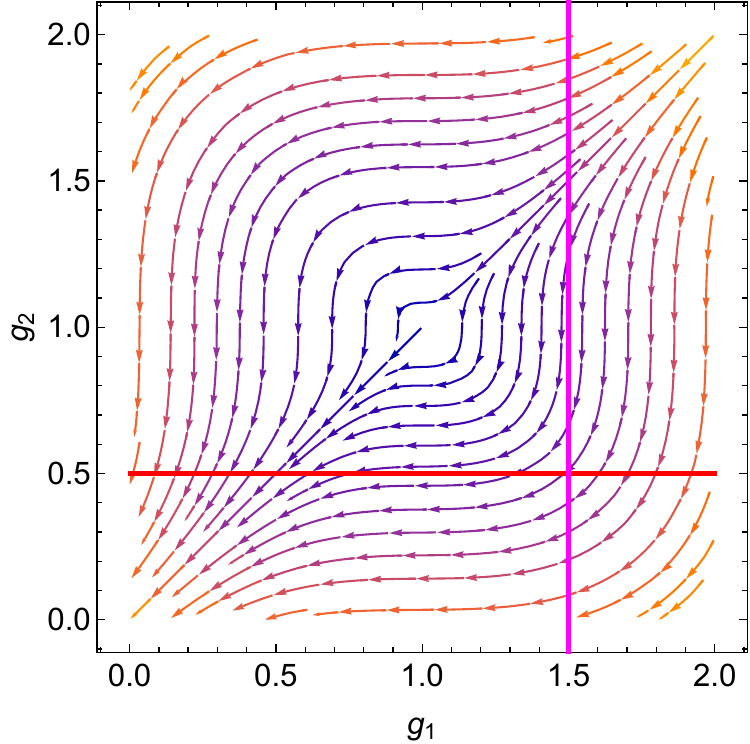}
\caption{\label{fig:shielding} We show a setup for a BVP with a non-degenerate fixed point (left panel) and a degenerate fixed point (right panel). 
The beta functions for the left panel are
$\beta_{g_1}=(g_2 - 1)$, $\beta_{g_2}= (g_1-1)$.
The beta functions for the right panel are $\beta_{g_1}=(g_2-1)^2$, $\beta_{g_2}= (g_1 - 1)^2$.
In the non-degenerate case, the fixed point shields the trajectories from reaching the IR boundary, whereas the degenerate case allows trajectories to pass through and, from the overall topology of the RG flow, one may already expect the BVP to not have a unique solution.}
\end{figure}

The several cases that we present now are inspired by the example above. The salient features of the system in Eqs.~\eqref{eq:beta1} and \eqref{eq:beta2} when it comes to the non-uniqueness of the boundary value problem are the following:
\begin{itemize}
\item[a)] Inside the region in the space of couplings in between the UV boundary and the IR boundary, there should be a smaller, bounded region that can be ``circumnavigated'' in two different ways. This can be achieved by a degenerate fixed point that lies within this region, i.e., if the fixed point corresponds to a multiple zero of the beta function. For non-degenerate fixed points, the direction of the flow on opposite sides of the fixed points is opposite -- one is either repelled or attracted by the fixed point. Therefore, a non-degenerate fixed point generally shields the IR boundary from the trajectories starting out at the UV boundary, cf.~Fig.~\ref{fig:shielding}. Accordingly, we require a degenerate fixed point. 
\item[b)] The RG trajectories should ``curve'' around this region, which requires that velocities of RG trajectories change direction during the flow. This requires the beta functions to contain higher-order terms in the couplings in which the two couplings mix, such as in the example in Eqs.~\eqref{eq:beta1} and \eqref{eq:beta2}. In terms of critical exponents of the fixed point, this is encoded in at least one complex critical exponent. Due to the degeneracy, one critical exponent is zero. 
\end{itemize}
The above features can be realized in three subcases, which are all deformations of one another:
\begin{itemize}
\item[1)] a degenerate fixed point,
\item[2)] two fixed points, which merge into a degenerate one under a suitable deformation of the beta functions,
\item[3)] no real fixed point, but a pair of complex conjugate fixed points, which merge into a degenerate, real fixed point under a suitable deformation of the beta functions.
\end{itemize}

\begin{figure}[!t]
\includegraphics[width=0.3\linewidth]{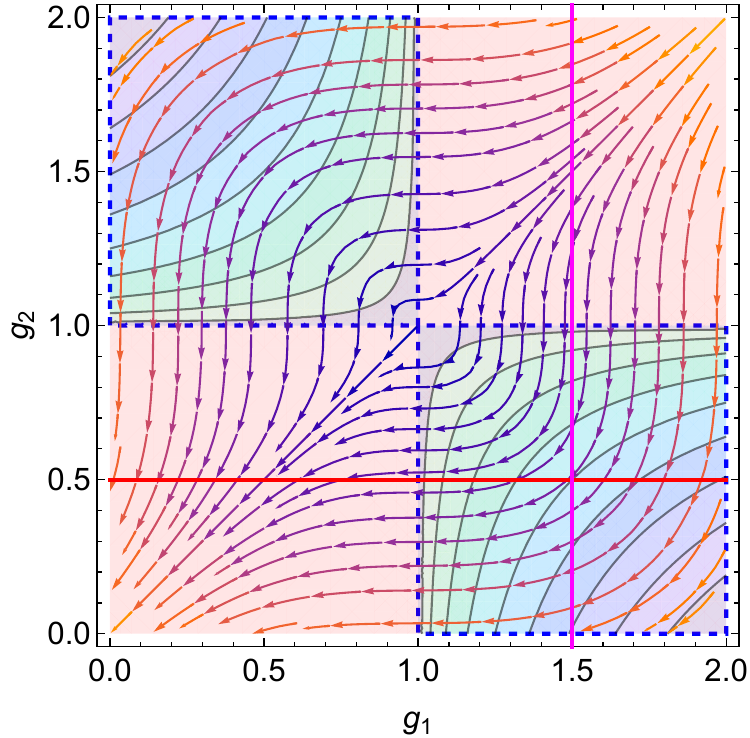}\includegraphics[width=0.3\linewidth]{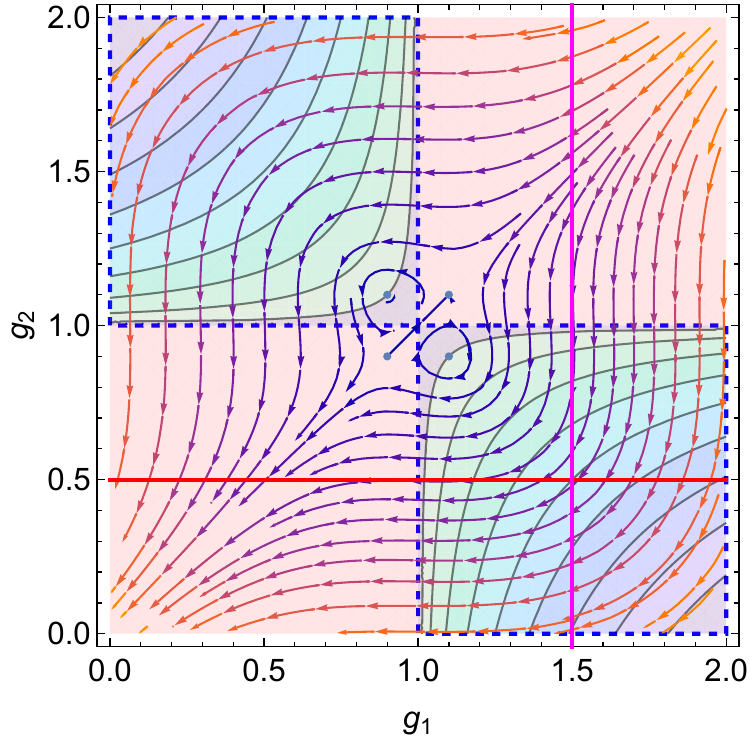}\includegraphics[width=0.36\linewidth]{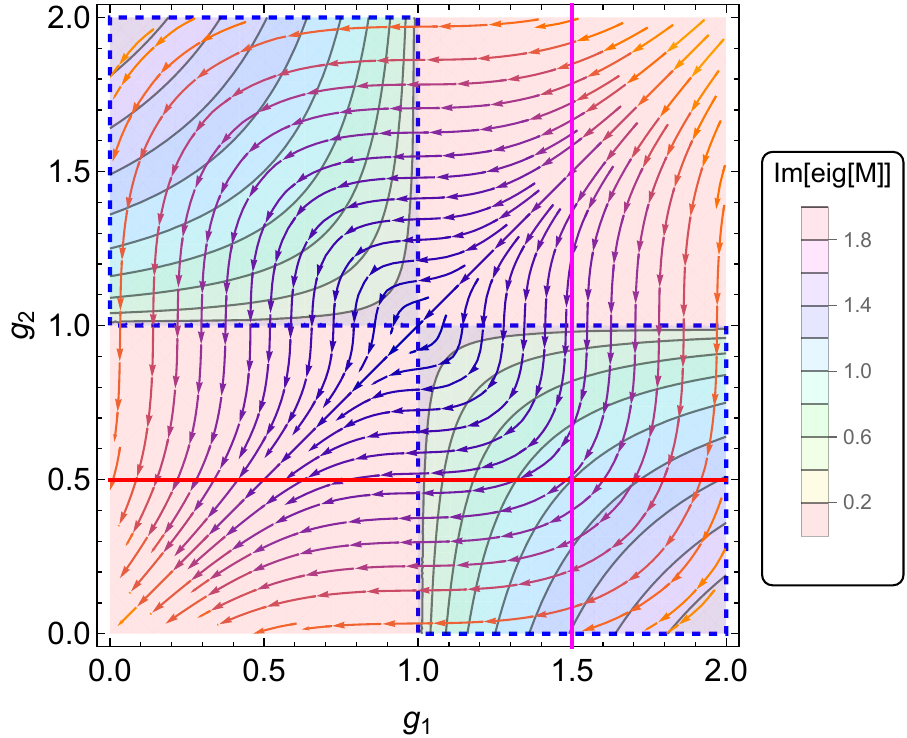}
\caption{\label{fig:streamplots_deg_FP} We show the RG flow generated by Eq.~\eqref{eq:beta1_deg} and \eqref{eq:beta2_deg} for the cases $a_{1,2}=1$, $b_{1,2}=1$, $c=0$ (left panel), $a_{1,2}=1\pm 0.1$, $b_{1,2}=1\pm 0.1$, $c=0$ (middle panel) and $a_{1,2}=1\pm 0.1$, $b_{1,2}=1\pm 0.1$, $c=0.0125$ (right panel). The left panel has a degenerate fixed point (with a four-fold degeneracy); the central panel has four real fixed points and the right panel no real fixed points. We also indicate the region where the eigenvalues of $\mathcal{J}$ are complex: These regions lie within the dashed blue boundaries, and the size of the imaginary part is indicated by the colored contours.
The UV boundary is indicated by the magenta line and the IR boundary by the red line. }
\end{figure}

Below, we illustrate these three cases and then show that the respective BVP does not have a unique solution. We highlight that in all cases, the fixed points (both real and complex) have complex critical exponents. Away from the fixed points, we calculate the eigenvalues of the Jacobian matrix $\mathcal{J}$. Evaluated on the fixed points, these become the critical exponents (times a negative sign). Since the beta functions are continuous functions of the couplings, by continuity, the eigenvalues of the Jacobian matrix $\mathcal{J}$ are complex in a region around the fixed point.
Our illustrative example is
\begin{eqnarray}
\beta_{g_1}&=&(g_2-a_1)(g_2-a_2)+c,\label{eq:beta1_deg}\\
\beta_{g_2}&=&(g_1-b_1)(g_1-b_2)+c.\label{eq:beta2_deg}
\end{eqnarray}
This results in 
\begin{equation}
{\rm eig}\left(\mathcal{J} \right)= \pm \sqrt{b_1+b_2-2g_1}\sqrt{a_1+a_2-2g_2},
\end{equation}
such that, for any choice of the constants, there are always values of the couplings such that the eigenvalues become complex. We illustrate this in Fig.~\ref{fig:streamplots_deg_FP}. In all cases, the region with complex Jacobian eigenvalues lies within the region traversed by the RG flow between the two boundaries.

\begin{figure}[!t]
\begin{center}
\includegraphics[width=0.6\linewidth]{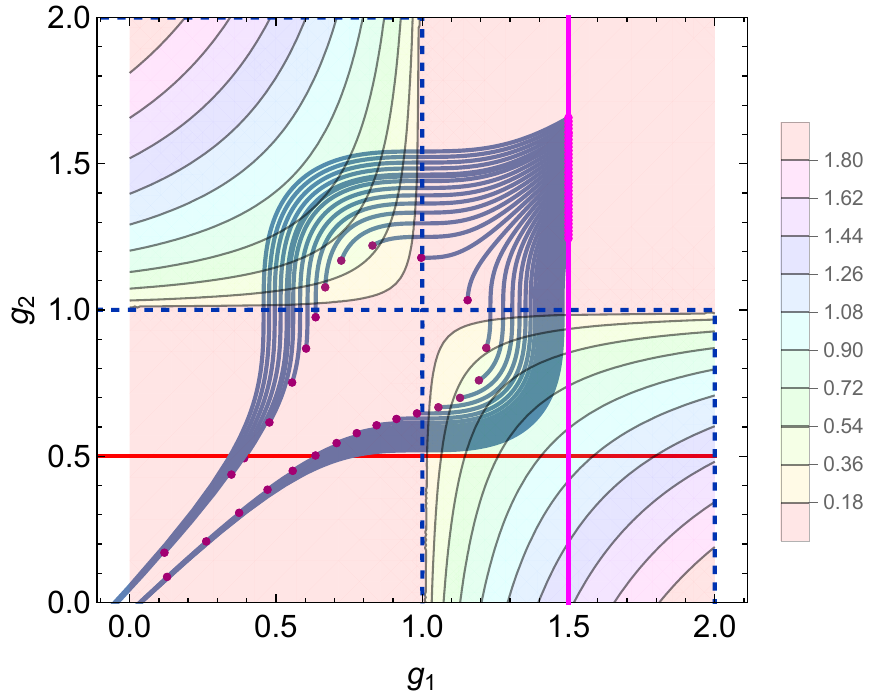}
\caption{\label{fig:onion_on_rainbow} We show trajectories that solve the RG flow for the case of a degenerate fixed point, described by Eq.~\eqref{eq:beta1_deg} and Eq.~\eqref{eq:beta2_deg} with $a_{1,2}=1$, $b_{1,2}=1$ and $c=0$. In the background, the colored contours indicate the size of the imaginary part of the eigenvalues of $\mathcal{J}$. The corresponding plots for the other two cases are very similar and thus we do not show them here. The endpoints of the RG trajectories, reached after a fixed amount of RG ``time'', $t = t_{\rm UV}- t_{\rm IR}$, are shown in dark red. Connecting them to a continuous curve would clearly give rise to the points $A$, $B$ and $C$ discussed in Sec.~\ref{sec:construction}.}
\end{center}
\end{figure}

The RG trajectories curve once they enter the region with nonvanishing imaginary parts of the eigenvalues of $\mathcal{J}$. Using ``left-turning'' and ``right-turning'' trajectories that circumnavigate the central region, we generate non-unique solutions to the BVP. Their existence is independent of the details of the central region, i.e., whether or not the fixed point is real and degenerate (with even multiplicity), real and non-degenerate, or complex. In all cases, we obtain sets of trajectories as illustrated in Fig.~\ref{fig:onion_on_rainbow}.

\begin{figure}[!t]
\includegraphics[width=0.45\linewidth]{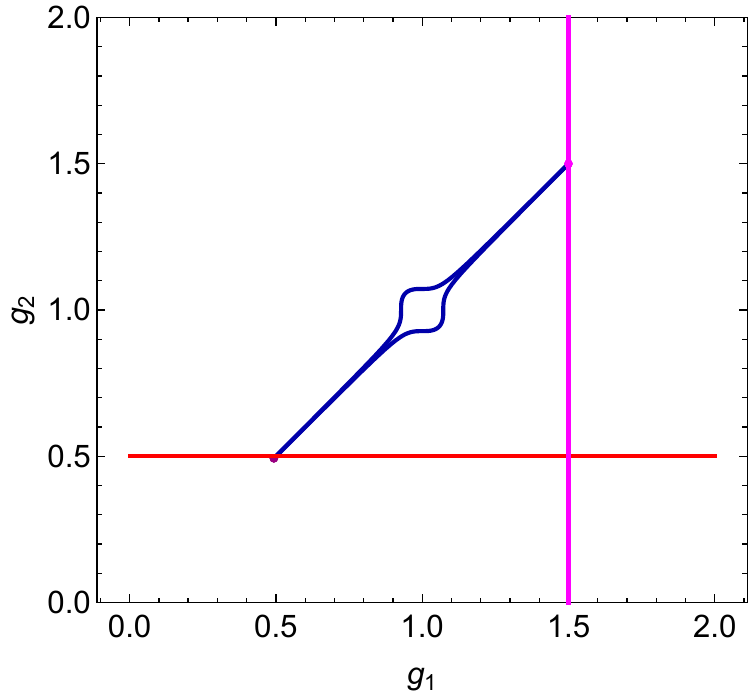}\quad\includegraphics[width=0.45\linewidth]{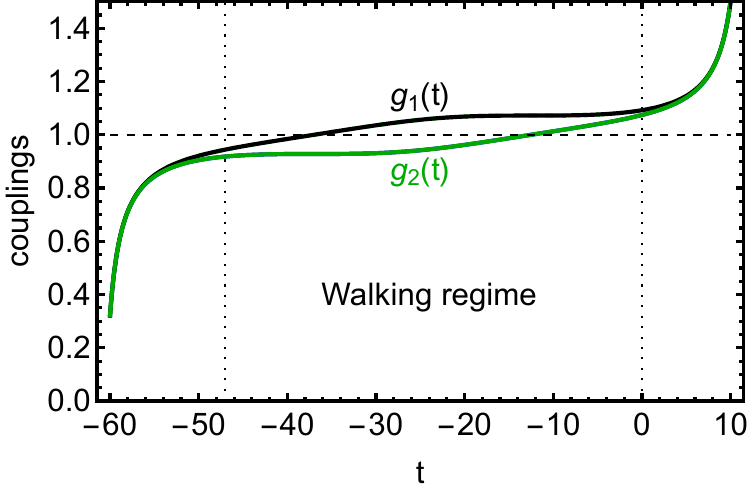}
\caption{\label{fig:walking_non_uniqueness} We show the two trajectories that constitute a pair of non-unique solutions to the
BVP for the RG flow described by Eq.~\eqref{eq:beta1_deg} and Eq.~\eqref{eq:beta2_deg} with $a_{1,2}=1$, $b_{1,2}=1$ and $c=0$ (left panel) and the couplings as a function of RG ``time'' (right panel). In the right panel, one can clearly see the walking regime once the couplings are close to the real part of the complex fixed point (horizontal dashed line).}
\end{figure}

We mention in passing that for the case in which the fixed points are complex, there is a range of choices for $c>0$ such that there are pairs of non-unique solutions to the BVP that exhibit walking \cite{Gorbenko:2018ncu}, see Fig.~\ref{fig:walking_non_uniqueness}.

%%%%%%
\subsection{Case II: No fixed point}
%%%%%%
While RG fixed points occur in many systems, they may still appear somewhat special -- after all, not every set of beta functions has a fixed point in any given region of coupling space. For polynomial beta functions, (complex) fixed points necessarily exist, but they may do so far from the region of interest in coupling space.
This makes it relevant to understand whether the origin of non-uniqueness of the BVP is tied to a fixed point in the corresponding region of the space of couplings or whether it merely relies on complex eigenvalues of the Jacobian matrix, even if these are not a consequence of continuity away from a fixed point with complex critical exponents.

As we have indicated above, a crucial property of the system of beta functions, enabling the non-uniqueness of the BVP, is that the Jacobian $\mathcal{J}$ has complex eigenvalues. This feature can also be achieved in settings in which there is no real fixed point within the region in between the UV boundary and the IR boundary. Next, we construct examples to illustrate this.

\begin{figure}[!t]
\includegraphics[width=0.45\linewidth]{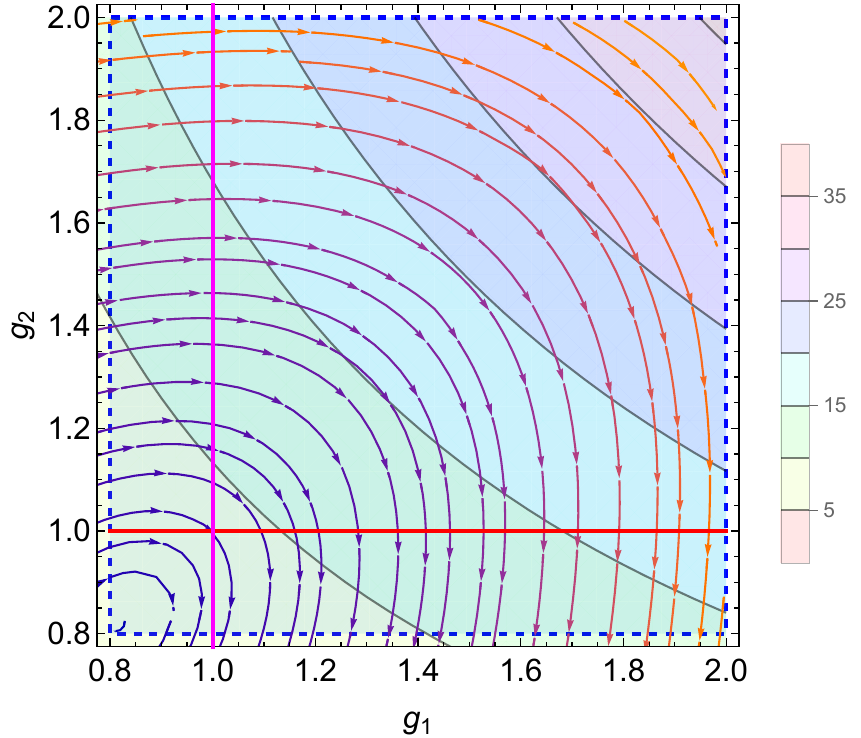}\quad\includegraphics[width=0.45\linewidth]{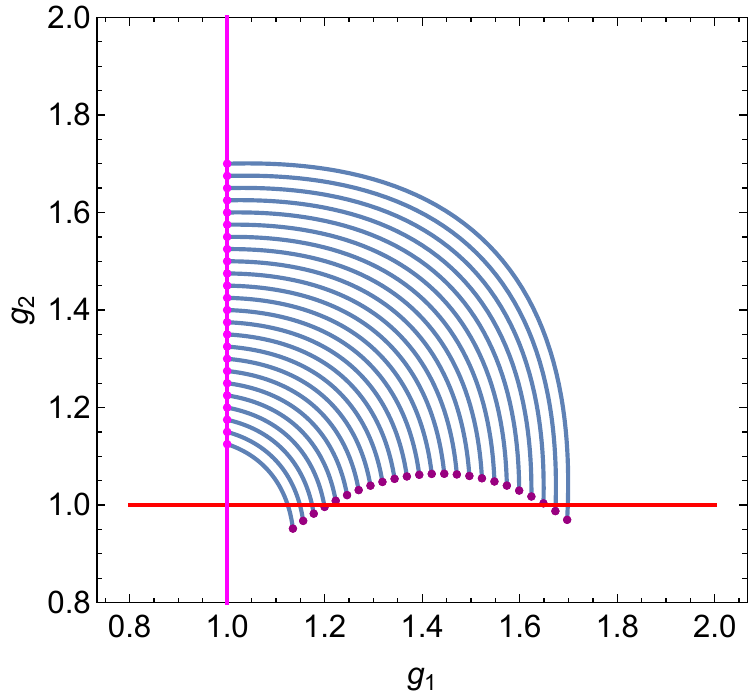}
\caption{\label{fig:FPoutsideregion} We show the RG flow for the system defined by Eq.~\eqref{eq:betag1noFPinregion} and Eq.~\eqref{eq:betag2noFPinregion}. We also indicate the size of the imaginary part of the eigenvalues of $\mathcal{J}$ through the colors. Finally, we show a set of trajectories that start at the UV boundary condition (magenta line) and evolve for a fixed amount of RG time. There are two such trajectories that end on the red line (IR boundary condition).}
\end{figure}

Such behavior can be achieved with a fixed point close to the intersection point of the UV boundary and the IR boundary, but outside the in-between region. 
As an example, we consider
\begin{eqnarray}
\beta_{g_1}&=&2g_1+10^{-2}g_1^2- 3g_2^3,\label{eq:betag1noFPinregion}\\
\beta_{g_2}&=&-2g_2 +3 g_1^3.\label{eq:betag2noFPinregion}
\end{eqnarray}
This gives rise to the RG flow and non-unique trajectories solving the BVP in Fig.~\ref{fig:FPoutsideregion}. The region in between the UV and the IR boundary does not contain a fixed point, but is characterized by imaginary parts in the eigenvalues of $\mathcal{J}$. This happens because the fixed point that lies just outside the region (at $g_1 \approx 0.82,\, g_2 \approx 0.82$) has complex critical exponents. By continuity of the beta functions, the eigenvalues of $\mathcal{J}$ away from the fixed point are therefore also complex. This results in the curving of trajectories that causes the points $A$, $B$ and $C$, described in Sec.~\ref{sec:construction}, to exist. 

\begin{figure}[!t]
\includegraphics[height=0.45\linewidth]{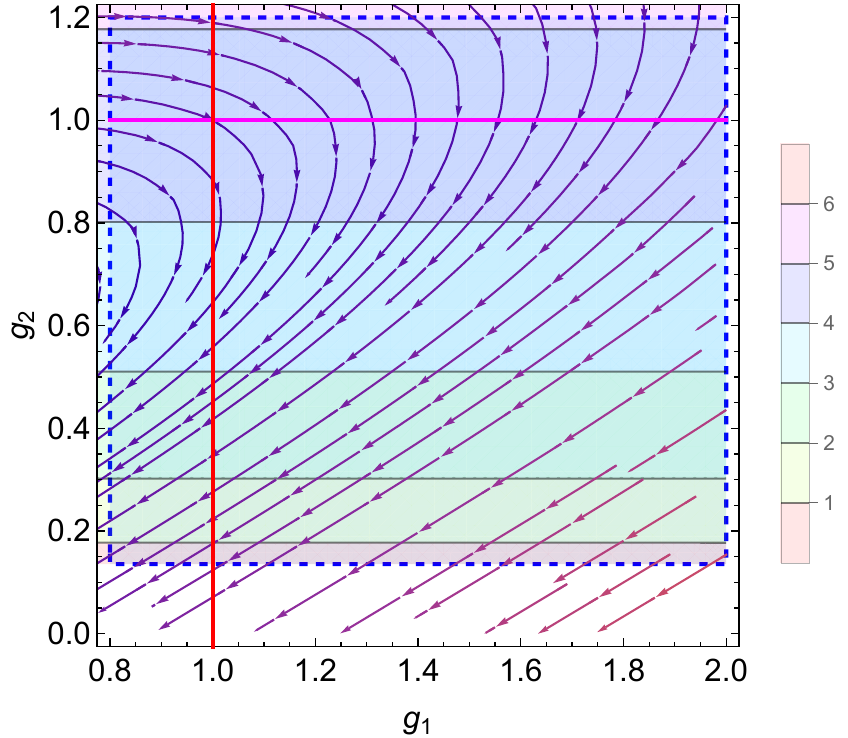}\quad 
\includegraphics[height=0.45\linewidth]{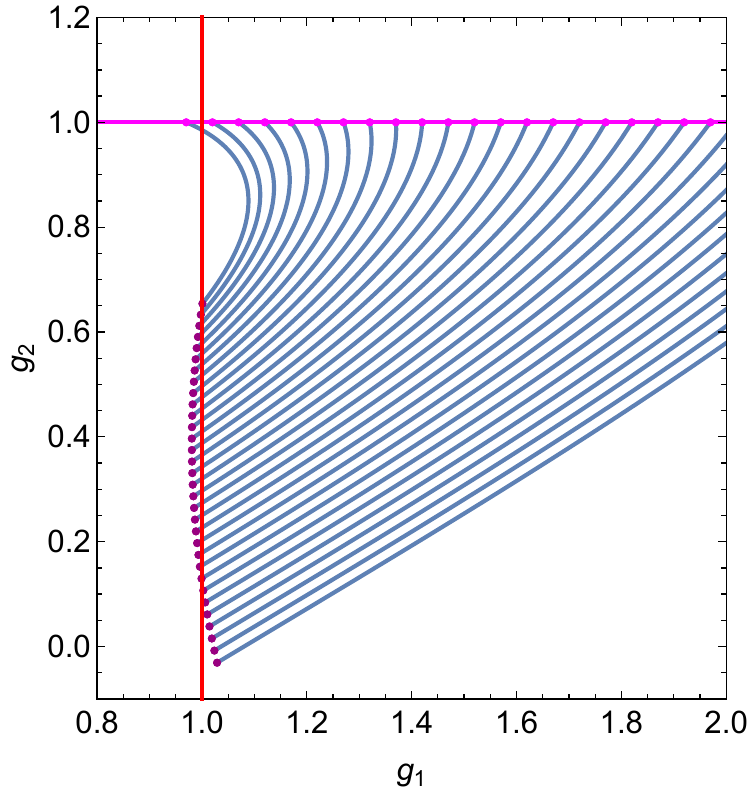}
\caption{\label{fig:complexM} We show the RG flow of the system defined by Eq.~\eqref{eq:beta1FPrealtheta} and Eq.~\eqref{eq:beta2FPrealtheta}. The eigenvalues of $\mathcal{J}$ are complex within the region surrounded by the blue dashed boundary; the magnitude of the imaginary part is indicated by the colors. We also show trajectories starting at the UV boundary (magenta line) that are integrated for a fixed amount of RG ``time''. Their endpoints intersect the IR boundary line twice, so there are two solutions to the BVP.}
\end{figure}

Finally, we construct a case in which the fixed points of beta functions lie outside of the region of interest, not close to this region, and have real critical exponents. Within the region between the UV boundary and the IR boundary, the eigenvalues of $\mathcal{J}$ are complex, and we find non-uniqueness, cf.~Fig.~\ref{fig:complexM}. 

The corresponding beta functions are given by
\begin{eqnarray}
\beta_{g_1} &=& 5g_1 -3g_2 - 4g_2^2,\label{eq:beta1FPrealtheta}\\
\beta_{g_2}&=&-2g_2 +3g_1.\label{eq:beta2FPrealtheta}
\end{eqnarray}

We caution that complex eigenvalues of $\mathcal{J}$ of course do not make an arbitrary BVP non-unique, but the boundaries have to be chosen judiciously. We also stress that we have presented a collection of examples suggesting a link between complex eigenvalues of $\mathcal{J}$ and non-uniqueness of the BVP. We are not aware of a proof of this link and providing one is certainly beyond the scope of this paper.

%%%%%%%%%%%%%%%%%%%%%%%%%%%%%%%%%%%%%%%%%%%%%%%%%%%%%%%%%

\section{A diagnostic for non-uniqueness}
\label{sec:SM}
%%%%%%%%%%%%%%%%%%%%%%%%%%%%%%%%%%%%%%%%%%%%%%%%%%%%%%%%%

In two-dimensional examples, the non-uniqueness of the BVP can be diagnosed straightforwardly from the flow diagrams.
In the above examples, e.g., the right panel in Fig.~\ref{fig:complexM}, the image of the UV hypersurface consists of
trajectories of fixed RG-time intervals.  A different diagnostic\footnote{Yet another diagnostic for non-uniqueness may be provided within degree theory \cite{Cronin1964,degreetheorybook}, although a straightforward application of the degree only provides information on the existence, not the uniqueness of solutions.} arises
by calculating the total RG-time to reach the IR hypersurface, for a sufficiently dense set of 
trajectories originating from the UV hypersurface. 
We consider the BVP defined by
\begin{equation}
g_1(k_{\rm IR})=a, \quad g_2(k_{\rm UV}) =b.
\end{equation}

For such a BVP, while keeping the boundaries fixed, we vary $g_1(k_{\rm UV})$, for fixed $k_{\rm UV}$. As a function of this parameter, the RG time it takes a trajectory to reach the IR boundary varies. We define the RG time interval as
\begin{equation}
{\Delta \rm T}_{\rm RG} = t_{\rm UV} - t_{\rm IR}(g_1(k_{\rm UV})),
\end{equation}
where $t= \ln k/k_0$, with a reference scale $k_0$. Herein, $t_{\rm IR}$ is a function of the initial condition $g_{1}(k_{\rm UV})$ and is the unique root of the equation
\begin{equation}
g_1(t_{\rm IR}) = a,
\end{equation}
where $g_1(t_{\rm IR})$ is considered a function of $t_{\rm IR}$ with the parameter $g_1(k_{\rm UV})$ fixed.

% \zois{In this way, given a two-dimensional \sout{boundary value problem} {\colas BVP} with boundary conditions
% \begin{eqnarray}
%     g_1(k_{\rm IR}) = a \;\;,\;\; g_2(k_{\rm UV}) = b \;,
% \end{eqnarray}
% one can define the total RG-time for all trajectories to reach the IR hypersurface (line) $g_1(k_{\rm UV}) = a$ as a function of $s = g_1(k_{\rm IR})$. We denote the solutions of the corresponding IVP for some $s$ as $(g_1(t;s), g_2(t;s))$. Then the RG-time interval function\sout{s} is given by\sout{,}
% \begin{eqnarray}
%     F(s) = t_0(s) - k_{\rm IR}\;\;,\;\;
% \end{eqnarray}
% where $t_0(s)$ is the (unique) root of the equation 
% \begin{eqnarray}
%     g_1(t;s) - g_1(k_{\rm IR}) = 0\;.
% \end{eqnarray}
% }

\begin{figure}[!t]
    \centering
    \includegraphics[%width=0.49\linewidth, 
    height=0.31\linewidth ]{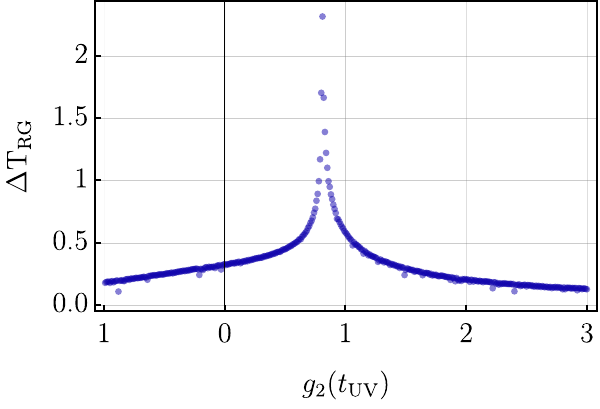}
    \quad
    \includegraphics[%width=0.49\linewidth
    height=0.31\linewidth
    ]{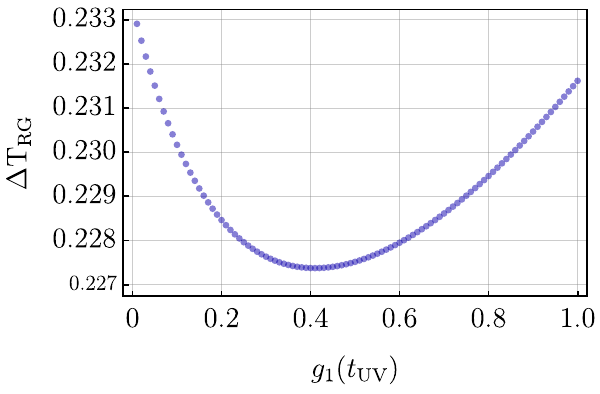}
    \caption{The RG-time interval function $\Delta T_{\rm RG}$ for the boundary value problems set up in Eqs.~\eqref{eq:beta1}-\eqref{eq:beta2} and in Eqs.~\eqref{eq:beta1FPrealtheta}- \eqref{eq:beta2FPrealtheta}. In both cases, the RG-time interval is non-monotonic. From the plots we already see that fixing the RG time we can find two initial conditions and thus two distinct solutions to the boundary value problem.}
    \label{fig:RGtimeplots}
\end{figure}

Consequently, if a BVP has non-unique solutions, ${\Delta \rm T}_{\rm RG}$ is necessarily non-monotonic. We demonstrate this diagnostic in Fig.~\eqref{fig:RGtimeplots} for the examples of \textbf{Case I} and \textbf{Case II}, since the existence, or not, of a fixed point changes the behaviour of  ${\Delta \rm T}_{\rm RG}$. The RG-time interval functions were calculated numerically. In the presence of a fixed point,  ${\Delta \rm T}_{\rm RG}$ has a singular point.
The RG trajectory corresponding to this singularity is the unique trajectory that connects the UV hypersurface with the fixed point. In the case where no fixed point is present in the integrating region the corresponding function is continuous and bounded, albeit still non-monotonic.

\subsection{A higher-dimensional example: Two-generation Gauge-Yukawa system}
\label{sec:gauge-Yukawa-example}

In higher-dimensional non-linear systems, it is significantly less efficient to apply the above diagnostic, because in general the IR endpoints of trajectories starting from a given UV boundary condition form a high-dimensional space, of which only a very particular subspace is actually of interest for a given BVP.  
This makes the numerical determination of the RG-time function less practical, because it would require computing a large number of trajectories as interpolation functions, of which only a tiny subset is ultimately retained. As an alternative, we provide a more generally applicable, albeit statistical, test for the non-uniqueness of higher-dimensional BVPs.\\
This test proceeds in three steps:
\begin{enumerate}
\item[1)] Set the UV initial conditions and sample the remaining, unfixed couplings in the UV with flat distributions within a given interval, e.g., bounded by perturbativity constraints. Run the RG flow to the IR for the resulting sets of initial conditions.
\item[2)] The resulting IR values of all couplings follow a distribution determined by step 1). From this distribution, select those trajectories, for which the IR values satisfy the IR boundary condition (with some desired precision). In practice, this means selecting all trajectories that end up in a bin in a histogram that shows the IR distribution of coupling values.
\item[3)] The trajectories selected in step 2 select a subset of coupling values in the UV from the original, flat distribution. If this new UV distribution is multimodal, i.e., contains sufficiently well-separated intervals in which it is nonzero and zero in between, then the BVP has as many solutions as there are disjoint intervals.
\end{enumerate}

As an example, we use the two-generation gauge-quark subsector of the Standard Model. 
The corresponding particle content is given by the gauge bosons of the U(1), SU(2) and SU(3) gauge groups, as well as the four Dirac fermions corresponding to the top, bottom, charm and strange quark.

In terms of couplings, this sector
consists of the three gauge couplings, four Yukawa couplings for the two generations of
quarks, i.e., top/bottom and charm/strange, as well as a mixing parameter $W$ between these two generations. The system of beta functions can be found in Eqs.~[35] and Eqs.~[38-39] in \cite{Alkofer:2020vtb}. 

 \begin{figure}[!t]
     \centering
  \includegraphics[width=0.8\linewidth]{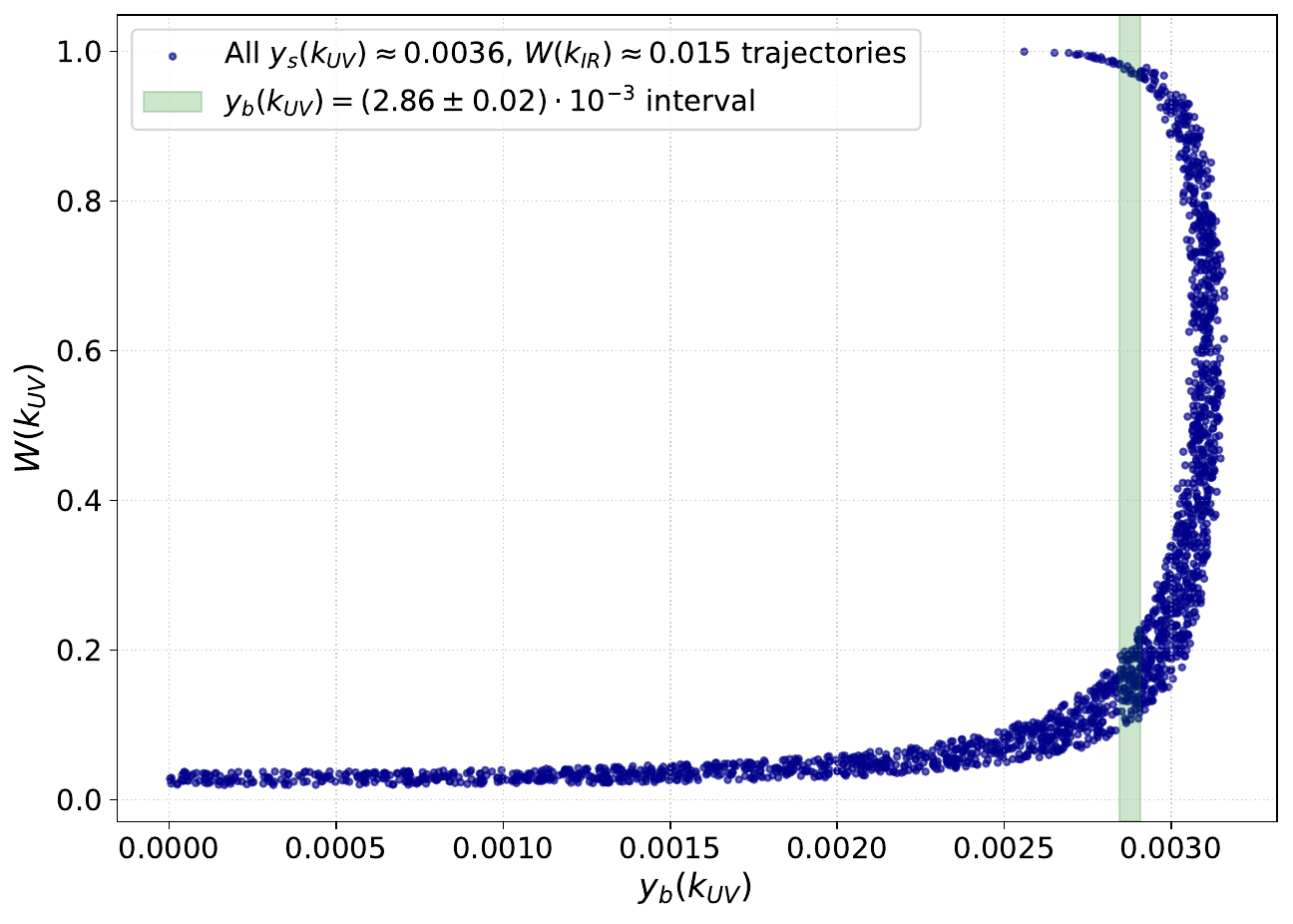}
     \caption{The points in the projected parameter space correspond to all initial conditions $(y_b,W)$ in the UV that give rise to trajectories with $W(k_{\rm IR}) \approx 0.01 -0.02$. These points form a hook that intersects the green region, corresponding to fixing the last $y_b(k_{\rm UV})$ condition, at two disconnected regions. In this case, the boundary value problem has two distinct solutions given that $y_b = (2.86 \pm 0.02)\cdot 10^{-3}$.} 
     \label{fig:diagnosticplot}
 \end{figure}
 
For this system, we explore whether some BVP, to be specified, has a non-unique solution. To that end, we need to find appropriate boundary conditions, since in general not all sets of boundary conditions exhibit non-uniqueness. In the two-dimensional examples presented above, a suitable choice of boundaries was straightforward by inspection of the two-dimensional plots of the RG flow. In the present, higher-dimensional parameter space, this is less straightforward. We therefore proceed by setting up \emph{multiple} BVPs simultaneously to discover whether there are BVPs which are non-unique among this larger set.\\
To this end, we set UV boundary conditions (at the Planck scale) for all gauge and Yukawa couplings, while the mixing parameter $W$ has a boundary condition at
the IR scale (the electroweak scale). To find whether there is \emph{some} choice of boundary conditions for which the BVP is non-unique, we consider \emph{test intervals} of UV initial conditions for some of the couplings with UV boundary conditions.
Specifically, we define such test intervals for $y_b(k_{\rm UV})$ and $y_s(k_{\rm UV})$. This choice is motivated by the fact that
the top quark is much heavier than all the other quarks, thus the flow of its Yukawa coupling effectively decouples from the other Yukawa couplings. Additionally, the top Yukawa coupling dominates the flow of the charm Yukawa coupling, so that its flow 
also decouples from that of the strange and bottom Yukawa couplings.
In contrast, the flows of $y_b$, $y_s$ and $W$ cannot be decoupled from each other, due to the strong dependence of $\beta_W$ on $y_b^2 -y_s^2$. 

Thus, we work with the following UV boundary conditions\footnote{Here we work with values of Yukawa couplings larger than the phenomenological ones, because then the RG flow from the Planck scale to the electroweak scale produces more pronounced structures. We checked explicitly that our qualitative results about non-uniqueness carry over to phenomenological values of top and charm quark Yukawa couplings.}
\begin{eqnarray}
    y_t(k_{\rm Planck}) &=& 9.2437\cdot 10^{-1}\,, \quad
    y_c(k_{\rm Planck}) = 3.4150\cdot 10^{-3},
\end{eqnarray}
and UV test intervals from which we ultimately select the UV boundary conditions. These larger test intervals are given by 
\begin{eqnarray}
0 \lesssim y_s(k_{\rm Planck}),y_b(k_{\rm Planck}) \lesssim 3.41\cdot 10^{-3}.
\end{eqnarray}

The test intervals for the bottom and strange quarks, as well as the larger top Yukawa coupling compared to its phenomenological value, are chosen in order to sample configurations where
\begin{eqnarray}\label{eq:strongflowcondtion}
   \frac{y_t^2}{16 \pi^2} \frac{y_s^2 + y_b^2}{y_s^2 -y_b^2} \gg 1\;.
\end{eqnarray}

This gives rise to a significant flow of the mixing element $W$ between the Planck scale and the electroweak scale, as one can see from the beta functions in \cite{Alkofer:2020vtb}. Finally, we work with an IR condition for $W$. Without the relation \eqref{eq:strongflowcondtion}, this BVP would have a unique solution simply because $W \approx \rm const$ would hold under the RG flow.

\begin{figure}[!t]
\begin{center}
\includegraphics[width=0.9\linewidth]{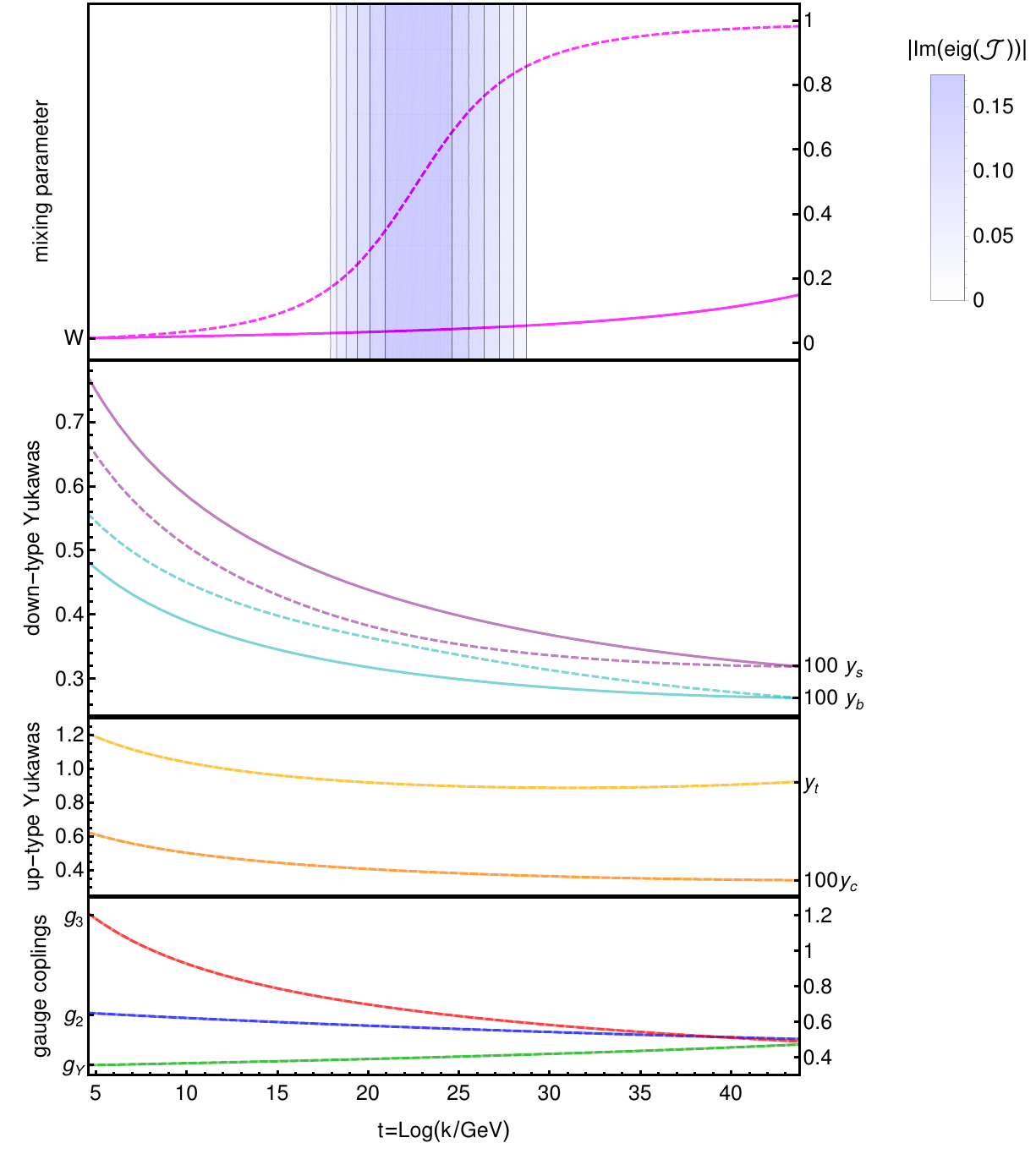}
\end{center}
\caption{\label{fig:eigJSM} 
We show two solutions to the BVP of the two-generation SM system as in~\cref{fig:diagnosticplot} for $y_b(k_\text{UV})=0.0027$ and $W(k_\text{IR})=0.015$ and all other couplings fixed to their SM values at $k_\text{UV}=M_\text{Planck}$. The two solutions correspond to $W(k_\text{UV})=0.15$ (continuous) and $W(k_\text{UV})=0.982$ (dashed). 
The upper panel shows $W$ (agreeing in the IR). 
In the background, we plot the imaginary part $\text{Im}(\text{eig}(\mathcal{J}))$ of the Jacobian eigenvalue (evaluated on the dashed solution) to indicate the region where it refocuses the trajectories and leads to non-uniqueness.
The other panels show the other evolved SM couplings to demonstrate that they agree in the UV.
}
\end{figure}

Because we flow from the UV to the IR, matching an exact IR boundary condition is challenging. Instead, we work with a narrow acceptance interval for the IR boundary condition, but have convinced ourselves that our results concerning non-uniqueness are not an artifact of this finite acceptance interval. We select
\begin{equation}
\; 0.01 \lesssim W(k_{\rm EW}) \lesssim 0.02\,\,.
\end{equation}
In practice, we sample the full UV interval
\begin{equation}
\; 0 \lesssim W(k_{\rm UV}) \lesssim 1\,\,,
\end{equation}
because we do not know a priori which UV value corresponds to an IR value within the acceptance interval. We proceed by numerically solving the flow equations for a set of $4\cdot 10^6$ initial conditions.  We then apply the statistical diagnostic described above.

 To do so, we fix $y_s(k_{\rm UV})= 3.364\cdot 10^{-3} \pm 1.7 \cdot 10^{-5}$ and plot all points that produce values in the IR acceptance interval for $W$ in the $W (k_{\rm UV})$ vs $y_b(k_{\rm UV})$ plane in Fig.~\eqref{fig:diagnosticplot}. We scan across the UV test interval for $y_b(k_{\rm UV})$ to determine whether there is a value of $y_b(k_{\rm UV})$ for which the distribution of $W(k_{\rm UV})$ is multimodal. We find that  
\begin{eqnarray}
       y_b(k_{\rm UV})= 2.86\cdot10^{-3}\pm 2 \cdot 10^{-5} \;,
\end{eqnarray}
 is such a value. Since we work numerically, we cannot fix the value with arbitrarily high precision, because this would require us to sample an arbitrarily high number of trajectories. For this value, there are two distinct initial conditions for $W(k_{\rm Planck})$, as shown in Fig.~\eqref{fig:diagnosticplot} and thus, two distinct solutions to the BVP.\footnote{We have convinced ourselves that the multimodality of the IR distribution is not an artifact of the finite precision with which we fix the values of couplings within our numerical study.} 

We provide an explicit example of a pair of trajectories satisfying the same set of boundary conditions. They differ in the UV value of the coupling $W$, for which an IR boundary condition is set, and in the IR values for some of those couplings for which a UV boundary condition is set. Non-uniqueness is again correlated with imaginary eigenvalues of $\mathcal{J}$ for one of the trajectories, cf.~Fig.~\ref{fig:eigJSM}. 

Although this pair of trajectories does not correspond to trajectories compatible with experiment, this example showcases that RG flows based on beta functions from systems with realistic field content can exhibit non-unique BVPs, in certain scenarios. 
This has implications, e.g., when it comes to accounting for the effect of new physics at a high scale, while fixing the values of some couplings at a low, experimentally accessible scale.
We comment on the implications of these results in the conclusions.

\subsection{Renormalization Group improvement in gravity: a cautionary example}
Asymptotically safe quantum gravity is an approach to quantum gravity based on a quantum field theory of the metric. Its UV regime is dominated by an RG fixed point. Accordingly, the study of RG flows takes center stage in this approach to quantum gravity. The ingredients for a possible non-uniqueness of a BVP also appear to be present: many calculations give rise to complex critical exponents of the interacting fixed point, e.g., \cite{Reuter:2001ag}, although cases with real-valued critical exponents also exist, e.g., \cite{Dona:2013qba}. 
The situation in which different couplings are fixed at different scales is realized in practical calculations, see, e.g., \cite{Reuter:2001ag,Gubitosi:2018gsl}. Most importantly, to fix the correct RG trajectory, one often uses RG improvement, which means that the physical scales at which experiments are conducted are equated with the RG scale $k$. For critical discussions of RG improvement in gravity, see \cite{Held:2021vwd,Knorr:2026vax}. Because measurements of the Newton coupling pertain to much smaller distance scales than measurements of the cosmological constant, in practice we have a BVP in which 
\begin{equation}
\frac{G(k_{\rm UV})}{k_{\rm UV}^2} =G_N,\quad\quad
\Lambda(k_{\rm IR})\cdot k_{\rm IR}^2 = \bar{\Lambda},
\end{equation}
where $G_N$ and $\bar{\Lambda}$ are the dimensionful couplings extracted from measurements, and $k_{\rm UV}= 10^{-5}\, \rm eV$ and $k_{\rm IR} = 10^{-33}\, \rm eV$, according to \cite{Gubitosi:2018gsl}. 

We take the complex critical exponents, together with these different RG scales at which couplings are typically fixed, as motivation to check whether the BVP has a unique solution.

\begin{figure}[!t]
\includegraphics[height=0.45\linewidth]{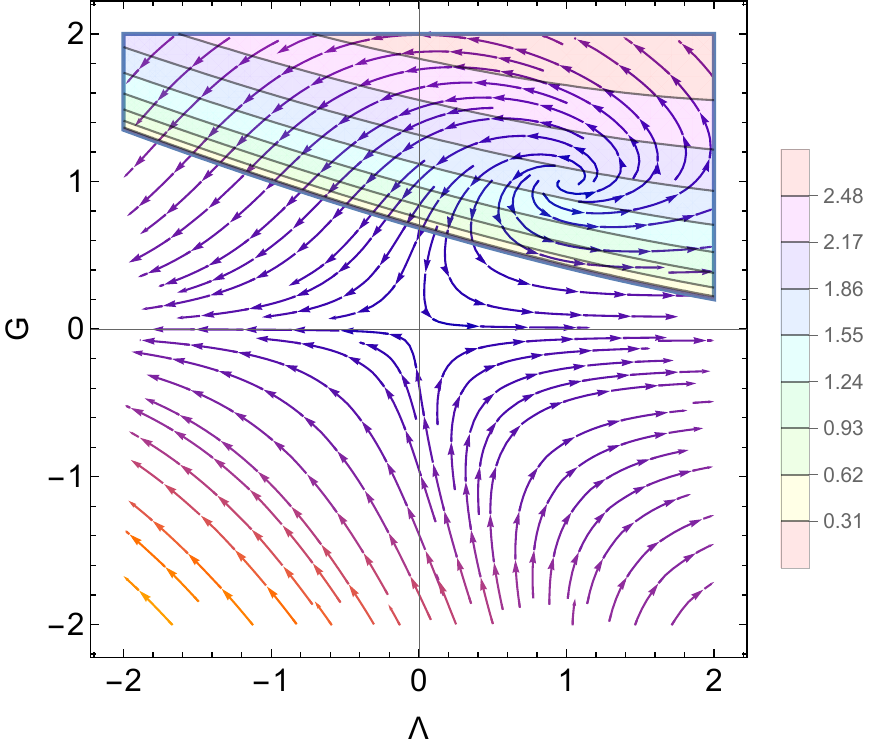}\quad \includegraphics[height=0.45\linewidth]{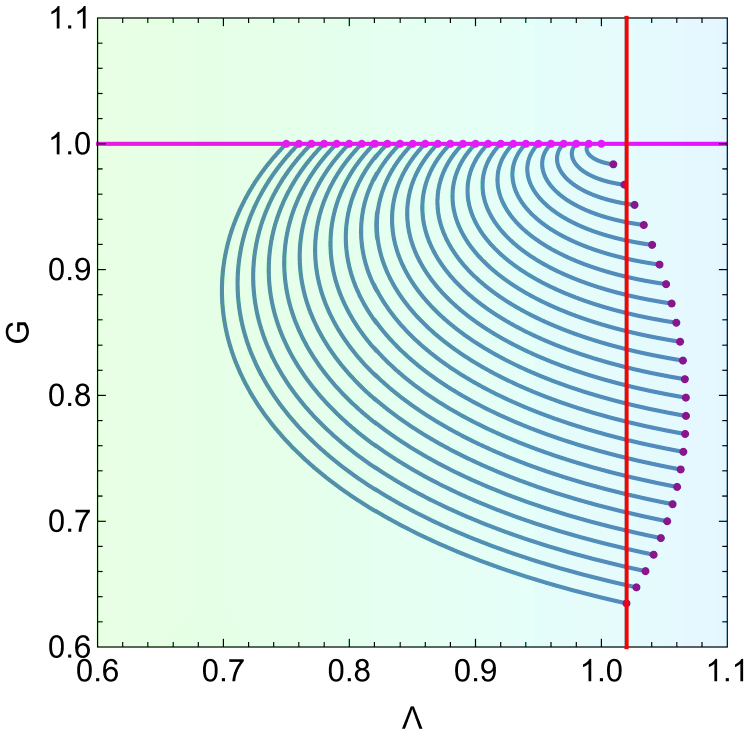}
\caption{\label{fig:EH} We show the RG flow according to the beta functions in Eqs.~\eqref{eq:EH1} and \eqref{eq:EH2} and an associated non-unique BVP.}
\end{figure}

We find complex eigenvalues of $\mathcal{J}$ in a region of the space of couplings, and, associated to that, non-uniqueness of a BVP, see Fig.~\ref{fig:EH}.

The beta functions we use are somewhat simplified from the typical beta functions for the Newton coupling $G$ and cosmological constant $\Lambda$ in the literature, but reproduce the RG flow as is very well known from asymptotically safe gravity, e.g., \cite{Reuter:2001ag} at a qualitative level.
We use
\begin{eqnarray}
\beta_G&=& 2 G - G^2 - G\cdot \Lambda,\label{eq:EH1}\\
\beta_{\Lambda}&=& -2 \Lambda + G \cdot \Lambda + G^2\label{eq:EH2},
\end{eqnarray}
with corresponding eigenvalues of the Jacobian matrix,
\begin{equation}
{\rm eig}(\mathcal{J})= \frac{1}{2}\left(-G-\Lambda \pm \sqrt{16 - 8 \Lambda + \Lambda^2-24 G + 2 G \cdot \Lambda + G^2} \right).
\end{equation}
The non-uniqueness of a suitably chosen BVP cautions against the use of RG improvement in gravity and calls for a determination of all couplings at a single scale $k \ll p_{\rm phys}$, with $p_{\rm phys}$ denoting the physical scales in the system.

We also add a positive note, namely that the BVP does have a unique solution in the region of coupling space that appears to be the phenomenologically relevant one, namely close to the Gaussian fixed point. There, the eigenvalues of $\mathcal{J}$ are real, and a BVP of the type described above, with $G(k_{\rm UV} =10^{-5}\, {\rm eV}) /k_{\rm UV}^2\approx 7 \cdot 10^{-57}\,{\rm eV}^{-2}$ and $\Lambda(k_{\rm IR} =10^{-33}\, {\rm eV})\cdot k_{\rm IR}^2\approx 10^{-66}{\rm eV}^2$, is actually unique.

We highlight that our results are achieved with a particular set of beta functions. It is well known that, when the full dynamics of asymptotic safety is restricted to a small subset of terms, such as the Einstein-Hilbert action, the beta functions acquire a dependence on unphysical choices (e.g., of the regulator function). This leads to variability of the beta functions in such small truncations, and beta functions both with and without complex critical exponents at the UV fixed point can be derived. We thus add the cautionary remark that the non-uniqueness that we observe may be an artifact of a small truncation, in which some choices of regulator function lead to complex critical exponents. In much larger truncations, such a property may potentially be absent, but this requires further study.

Our results highlight that, due to the fortunate coincidence that the measured values result in a unique solution, the possibility of a non-unique solution has not affected any studies in asymptotically safe gravity. Nevertheless, at a conceptual level, this may be taken as a cautionary result regarding the use of RG improvement in gravity, adding to the results in \cite{Held:2021vwd, Knorr:2026vax}.

%%%%
\section{Conclusions and outlook}\label{sec:conclusions}
%%%%
Motivated by settings in which constraints on different couplings arise at different scales, we consider BVPs for finite-dimensional RG flows. In such systems, the RG equations correspond to first-order ordinary differential equations. Within the RG flow, trajectories cannot cross. Therefore, supplying $N$ initial conditions for $N$ couplings at a UV scale $k_{\rm UV}$ results in a unique solution to the IVP, provided the beta functions are sufficiently smooth. RG flows within a given set of couplings are reversible. Thus, supplying $N$ initial conditions for $N$ couplings at an IR scale $k_{\rm IR}$ also leads to a unique solution.\footnote{Under RG flows towards the UV, irrelevant interactions increase their distance to a fixed point. Thus, for each irrelevant coupling, the initial condition in the IR must be supplied with very high precision.}
At first glance, one may thus be tempted to think that supplying $N-k$ conditions at the UV scale $k_{\rm UV}$ and $k$ conditions at the IR scale $k_{\rm IR}$ should also select a unique RG trajectory. However, this is not the case: uniqueness is not guaranteed.

The prerequisite for non-uniqueness to the BVP is the following: the RG-time duration changes non-monotonically across a set of neighboring trajectories that connect a hypersurface defined by UV boundary conditions to another hypersurface defined by IR boundary conditions. Therefore, within this set of trajectories, there are subsets (in the simplest case pairs) of trajectories which take the same amount of RG time to connect the two hypersurfaces.

Overall, we find that all of the investigated examples are consistent with the conclusion that non-uniqueness of a BVP occurs when the Jacobian matrix of the intermediate flow exhibits complex eigenvalues.
We are not aware of a mathematical theorem that confirms this observation.
\\

In settings with two couplings, determining whether or not a BVP is non-unique is relatively straightforward: For a given set of trajectories that share a UV boundary condition, one tests whether the line of endpoints (given by the values of the couplings reached after a fixed amount of RG time) intersects the line defined by the IR boundary condition more than once.

In settings with several couplings, one can set up a statistical test in order to find whether the BVP is non-unique, and if so, for which values of the UV and IR boundary conditions. To that end, we sample larger test intervals of the couplings and explore multiple BVPs simultaneously. To do so, we run a large number of RG trajectories from the UV to the IR. Then, we investigate the IR distribution of those couplings for which IR boundary conditions are imposed in each BVP separately. For those choices of boundary conditions for which the UV distribution is multimodal, the specific BVP is non-unique.

We demonstrate explicitly that this statistical test is applicable in a case with 8 couplings (three gauge couplings, four Yukawa couplings, one mixing parameter). Once we have thereby identified a non-unique BVP, we also find that the corresponding trajectories traverse a region in which the eigenvalues of the Jacobian matrix has a non-vanishing imaginary part.
This example also highlights that physically relevant systems such as the RG flow in a subsector of the Standard Model can exhibit non-unique BVPs.

In practice, this may be relevant when a setting with new 
physics at a high scale is confronted with observations. A subset of 
couplings may be fixed by the properties of the new physics 
at the high scale and another subset may be fixed by 
experimental data. A concrete example in which a subset of couplings is fixed by new physics at high scales is asymptotically safe quantum gravity \cite{Shaposhnikov:2009pv, Eichhorn:2017ylw, Eichhorn:2017lry,Eichhorn:2018whv, Eichhorn:2025sux}, where the high scale is the Planck scale. However, the new physics could also be sub-Planckian, and consist simply of additional matter fields beyond the SM, which, when integrated out, fix a subset of couplings from consistency conditions of the underlying UV theory. To the best of our knowledge, it has not yet been pointed out that such setups may be intrinsically ill-defined and it must be carefully investigated for each case whether the corresponding values of couplings lead to a unique or non-unique BVP. The diagnostic we have presented here can achieve this.
\\

For future research, testing our hypothesis that complex eigenvalues of the Jacobian matrix enable non-uniqueness of the BVP is clearly a relevant goal. 
These are linked to how trajectories ``curve'' in such a way that several trajectories connect a UV boundary surface with an IR boundary surface within the same amount of RG time. This is conceptually reminiscent of gravitational lensing in non-flat spacetimes, in which there can be more than one geodesic connecting two distinct spacetime points. Thus, formulating RG flows in the language of differential geometry and extracting a notion of curvature from beta functions may be a fruitful direction for future research.

Understanding whether there are other necessary or sufficient conditions for non-uniqueness is another relevant goal.
We have left aside the fact that beta functions are not invariant under reparameterizations of couplings. Only universal quantities, such as the critical exponents of fixed points, are invariant under reparameterizations. Investigating how the non-uniqueness of the BVP is affected by reparameterizations of couplings is another point for future studies.

In addition, complex eigenvalues of the Jacobian matrix can (although they do not need to) occur in conjunction with limit cycles. In turn, limit cycles, while not automatically excluded by the $a-$ and $c-$theorem, at least make the question whether there is a quantity that decreases monotonically under the RG flow more subtle. Thus, investigating the interplay between non-uniqueness of the BVP and the $a-$ and $c-$ theorem is a further question of interest.

More pragmatically, investigating the RG flows of concrete physical examples to establish which physical systems can exhibit such non-uniqueness is of interest. This is linked to the question for which systems it is actually viable to use information from several distinct scales in order to determine the RG trajectory.
This last question may even have implications for the interpretation of experimental results, if they correspond to measurements at distinct energy scales and are used to fix the values of distinct couplings at different scales. \newline\\

\noindent \textbf{\textit{Acknowledgments.}}
 We acknowledge the European Research Council's (ERC) support under the European Union’s Horizon 2020 research and innovation program Grant agreement No.~101170215 (ProbeQG).
This work is supported by the Deutsche Forschungsgemeinschaft (DFG, German Research Foundation) under Germany’s Excellence Strategy EXC 2181/1 - 390900948 (the Heidelberg STRUCTURES Excellence Cluster).

%=======================================================================================================

\newpage

\bibliography{References}

\providecommand{\href}[2]{#2}\begingroup\raggedright\begin{thebibliography}{10}

\bibitem{Buccio:2024hys}
D.~Buccio, J.F.~Donoghue, G.~Menezes and R.~Percacci, \emph{{Physical Running
  of Couplings in Quadratic Gravity}},
  \href{https://doi.org/10.1103/PhysRevLett.133.021604}{\emph{Phys. Rev. Lett.}
  {\bfseries 133} (2024) 021604}
  [\href{https://arxiv.org/abs/2403.02397}{{\ttfamily 2403.02397}}].

\bibitem{Buttazzo:2013uya}
D.~Buttazzo, G.~Degrassi, P.P.~Giardino, G.F.~Giudice, F.~Sala, A.~Salvio
  et~al., \emph{{Investigating the near-criticality of the Higgs boson}},
  \href{https://doi.org/10.1007/JHEP12(2013)089}{\emph{JHEP} {\bfseries 12}
  (2013) 089} [\href{https://arxiv.org/abs/1307.3536}{{\ttfamily 1307.3536}}].

\bibitem{Hambye:1996wb}
T.~Hambye and K.~Riesselmann, \emph{{Matching conditions and Higgs mass upper
  bounds revisited}},
  \href{https://doi.org/10.1103/PhysRevD.55.7255}{\emph{Phys. Rev. D}
  {\bfseries 55} (1997) 7255}
  [\href{https://arxiv.org/abs/hep-ph/9610272}{{\ttfamily hep-ph/9610272}}].

\bibitem{Bezrukov:2012sa}
F.~Bezrukov, M.Y.~Kalmykov, B.A.~Kniehl and M.~Shaposhnikov, \emph{{Higgs Boson
  Mass and New Physics}},
  \href{https://doi.org/10.1007/JHEP10(2012)140}{\emph{JHEP} {\bfseries 10}
  (2012) 140} [\href{https://arxiv.org/abs/1205.2893}{{\ttfamily 1205.2893}}].

\bibitem{ParticleDataGroup:2024cfk}
{\scshape Particle Data Group} collaboration, \emph{{Review of particle
  physics}}, \href{https://doi.org/10.1103/PhysRevD.110.030001}{\emph{Phys.
  Rev. D} {\bfseries 110} (2024) 030001}.

\bibitem{Buccio:2023lzo}
D.~Buccio, J.F.~Donoghue and R.~Percacci, \emph{{Amplitudes and renormalization
  group techniques: A case study}},
  \href{https://doi.org/10.1103/PhysRevD.109.045008}{\emph{Phys. Rev. D}
  {\bfseries 109} (2024) 045008}
  [\href{https://arxiv.org/abs/2307.00055}{{\ttfamily 2307.00055}}].

\bibitem{Eichhorn:2022jqj}
A.~Eichhorn, \emph{{Status update: Asymptotically safe gravity-matter
  systems}}, \href{https://doi.org/10.1393/ncc/i2022-22029-4}{\emph{Nuovo Cim.
  C} {\bfseries 45} (2022) 29}
  [\href{https://arxiv.org/abs/2201.11543}{{\ttfamily 2201.11543}}].

\bibitem{Eichhorn:2022gku}
A.~Eichhorn and M.~Schiffer, \emph{Asymptotic safety of gravity with matter},
  in \emph{Handbook of Quantum Gravity}, C.~Bambi, L.~Modesto and I.~Shapiro,
  eds., pp.~915--1001, Springer Nature Singapore (2024).

\bibitem{Shaposhnikov:2009pv}
M.~Shaposhnikov and C.~Wetterich, \emph{{Asymptotic safety of gravity and the
  Higgs boson mass}},
  \href{https://doi.org/10.1016/j.physletb.2009.12.022}{\emph{Phys. Lett. B}
  {\bfseries 683} (2010) 196}
  [\href{https://arxiv.org/abs/0912.0208}{{\ttfamily 0912.0208}}].

\bibitem{Eichhorn:2017ylw}
A.~Eichhorn and A.~Held, \emph{{Top mass from asymptotic safety}},
  \href{https://doi.org/10.1016/j.physletb.2017.12.040}{\emph{Phys. Lett. B}
  {\bfseries 777} (2018) 217}
  [\href{https://arxiv.org/abs/1707.01107}{{\ttfamily 1707.01107}}].

\bibitem{Eichhorn:2017lry}
A.~Eichhorn and F.~Versteegen, \emph{{Upper bound on the Abelian gauge coupling
  from asymptotic safety}},
  \href{https://doi.org/10.1007/JHEP01(2018)030}{\emph{JHEP} {\bfseries 01}
  (2018) 030} [\href{https://arxiv.org/abs/1709.07252}{{\ttfamily
  1709.07252}}].

\bibitem{Eichhorn:2018whv}
A.~Eichhorn and A.~Held, \emph{{Mass difference for charged quarks from
  asymptotically safe quantum gravity}},
  \href{https://doi.org/10.1103/PhysRevLett.121.151302}{\emph{Phys. Rev. Lett.}
  {\bfseries 121} (2018) 151302}
  [\href{https://arxiv.org/abs/1803.04027}{{\ttfamily 1803.04027}}].

\bibitem{Eichhorn:2025sux}
A.~Eichhorn, Z.~Gyftopoulos and A.~Held, \emph{{Quark and lepton mixing in the
  asymptotically safe Standard Model}},  7, 2025.

\bibitem{keller2018numerical}
H.~Keller, \emph{Numerical Methods for Two-Point Boundary-Value Problems},
  Dover Books on Mathematics, Dover Publications (2018).

\bibitem{Peterson1977}
A.C.~Peterson, \emph{Existence-uniqueness for two-point boundary value problems
  for $n$th order nonlinear differential equations},
  \href{https://doi.org/10.1216/RMJ-1977-7-1-103}{\emph{Rocky Mountain Journal
  of Mathematics} {\bfseries 7} (1977) 103}.

\bibitem{Agarwal1986_TwoPointProblems}
R.P.~Agarwal, \emph{Some new results on two-point problems for higher order
  differential equations}, {\emph{Funkcialaj Ekvacioj} {\bfseries 29} (1986)
  197}.

\bibitem{Bulycheva:2014twa}
K.M.~Bulycheva and A.S.~Gorsky, \emph{{Limit cycles in renormalization group
  dynamics}}, \href{https://doi.org/10.3367/UFNe.0184.201402g.0182}{\emph{Phys.
  Usp.} {\bfseries 57} (2014) 171}
  [\href{https://arxiv.org/abs/1402.2431}{{\ttfamily 1402.2431}}].

\bibitem{Jepsen:2020czw}
C.B.~Jepsen, I.R.~Klebanov and F.K.~Popov, \emph{{RG limit cycles and
  unconventional fixed points in perturbative QFT}},
  \href{https://doi.org/10.1103/PhysRevD.103.046015}{\emph{Phys. Rev. D}
  {\bfseries 103} (2021) 046015}
  [\href{https://arxiv.org/abs/2010.15133}{{\ttfamily 2010.15133}}].

\bibitem{LeClair:2002ux}
A.~LeClair, J.M.~Roman and G.~Sierra, \emph{{Russian doll renormalization group
  and superconductivity}},
  \href{https://doi.org/10.1103/PhysRevB.69.020505}{\emph{Phys. Rev. B}
  {\bfseries 69} (2004) 020505}
  [\href{https://arxiv.org/abs/cond-mat/0211338}{{\ttfamily
  cond-mat/0211338}}].

\bibitem{Braaten:2004rn}
E.~Braaten and H.W.~Hammer, \emph{{Universality in few-body systems with large
  scattering length}},
  \href{https://doi.org/10.1016/j.physrep.2006.03.001}{\emph{Phys. Rept.}
  {\bfseries 428} (2006) 259}
  [\href{https://arxiv.org/abs/cond-mat/0410417}{{\ttfamily
  cond-mat/0410417}}].

\bibitem{Zamolodchikov:1986gt}
A.B.~Zamolodchikov, \emph{{Irreversibility of the Flux of the Renormalization
  Group in a 2D Field Theory}}, {\emph{JETP Lett.} {\bfseries 43} (1986) 730}.

\bibitem{Cardy:1988cwa}
J.L.~Cardy, \emph{{Is There a c Theorem in Four-Dimensions?}},
  \href{https://doi.org/10.1016/0370-2693(88)90054-8}{\emph{Phys. Lett. B}
  {\bfseries 215} (1988) 749}.

\bibitem{Anselmi:1997am}
D.~Anselmi, D.Z.~Freedman, M.T.~Grisaru and A.A.~Johansen,
  \emph{{Nonperturbative formulas for central functions of supersymmetric gauge
  theories}}, \href{https://doi.org/10.1016/S0550-3213(98)00278-8}{\emph{Nucl.
  Phys. B} {\bfseries 526} (1998) 543}
  [\href{https://arxiv.org/abs/hep-th/9708042}{{\ttfamily hep-th/9708042}}].

\bibitem{Komargodski:2011vj}
Z.~Komargodski and A.~Schwimmer, \emph{{On Renormalization Group Flows in Four
  Dimensions}}, \href{https://doi.org/10.1007/JHEP12(2011)099}{\emph{JHEP}
  {\bfseries 12} (2011) 099} [\href{https://arxiv.org/abs/1107.3987}{{\ttfamily
  1107.3987}}].

\bibitem{Curtright:2011qg}
T.L.~Curtright, X.~Jin and C.K.~Zachos, \emph{{RG flows, cycles, and c-theorem
  folklore}}, \href{https://doi.org/10.1103/PhysRevLett.108.131601}{\emph{Phys.
  Rev. Lett.} {\bfseries 108} (2012) 131601}
  [\href{https://arxiv.org/abs/1111.2649}{{\ttfamily 1111.2649}}].

\bibitem{Gorbenko:2018ncu}
V.~Gorbenko, S.~Rychkov and B.~Zan, \emph{{Walking, Weak first-order
  transitions, and Complex CFTs}},
  \href{https://doi.org/10.1007/JHEP10(2018)108}{\emph{JHEP} {\bfseries 10}
  (2018) 108} [\href{https://arxiv.org/abs/1807.11512}{{\ttfamily
  1807.11512}}].

\bibitem{Cronin1964}
J.~Cronin, \emph{Fixed points and topological degree in nonlinear analysis},
  American Mathematical Society (1964),
  \href{https://doi.org/10.1201/9781420011487}{10.1201/9781420011487}.

\bibitem{degreetheorybook}
Y.~Cho and Y.-Q.~Chen, \emph{Topological Degree Theory and Applications (1st
  ed.)}, Chapman and Hall/CRC (2006),
  \href{https://doi.org/10.1201/9781420011487}{10.1201/9781420011487}.

\bibitem{Alkofer:2020vtb}
R.~Alkofer, A.~Eichhorn, A.~Held, C.M.~Nieto, R.~Percacci and M.~Schr{\"o}fl,
  \emph{{Quark masses and mixings in minimally parameterized UV completions of
  the Standard Model}},
  \href{https://doi.org/10.1016/j.aop.2020.168282}{\emph{Annals Phys.}
  {\bfseries 421} (2020) 168282}
  [\href{https://arxiv.org/abs/2003.08401}{{\ttfamily 2003.08401}}].

\bibitem{Reuter:2001ag}
M.~Reuter and F.~Saueressig, \emph{{Renormalization group flow of quantum
  gravity in the Einstein-Hilbert truncation}},
  \href{https://doi.org/10.1103/PhysRevD.65.065016}{\emph{Phys. Rev. D}
  {\bfseries 65} (2002) 065016}
  [\href{https://arxiv.org/abs/hep-th/0110054}{{\ttfamily hep-th/0110054}}].

\bibitem{Dona:2013qba}
P.~Don{\`a}, A.~Eichhorn and R.~Percacci, \emph{{Matter matters in
  asymptotically safe quantum gravity}},
  \href{https://doi.org/10.1103/PhysRevD.89.084035}{\emph{Phys. Rev. D}
  {\bfseries 89} (2014) 084035}
  [\href{https://arxiv.org/abs/1311.2898}{{\ttfamily 1311.2898}}].

\bibitem{Gubitosi:2018gsl}
G.~Gubitosi, R.~Ooijer, C.~Ripken and F.~Saueressig, \emph{{Consistent early
  and late time cosmology from the RG flow of gravity}},
  \href{https://doi.org/10.1088/1475-7516/2018/12/004}{\emph{JCAP} {\bfseries
  12} (2018) 004} [\href{https://arxiv.org/abs/1806.10147}{{\ttfamily
  1806.10147}}].

\bibitem{Held:2021vwd}
A.~Held, \emph{{Invariant Renormalization-Group improvement}},
  \href{https://arxiv.org/abs/2105.11458}{{\ttfamily 2105.11458}}.

\bibitem{Knorr:2026vax}
B.~Knorr, \emph{{Asymptotically (un)safe scattering amplitudes from scratch: a
  deep dive into the IR jungle}},
  \href{https://arxiv.org/abs/2602.21285}{{\ttfamily 2602.21285}}.

\end{thebibliography}\endgroup
    
\end{document}